\documentclass[twocolumn,twoside]{IEEEtran}

\ifCLASSINFOpdf
  
\else
 
\fi

\ifCLASSOPTIONcompsoc
\usepackage[caption=false, font=normalsize, labelfont=sf, textfont=sf]{subfig}
\else
\usepackage[caption=false, font=normalsize]{subfig}
\fi
\usepackage{lipsum}%
\usepackage[dvipsnames]{xcolor}

\usepackage{balance}
\usepackage{multicol}   
\usepackage{cite}
\usepackage{gensymb}
\usepackage{multirow}
\usepackage{graphics}
\usepackage{epsfig}
\usepackage{graphicx}
\usepackage{epstopdf}
\usepackage{textcomp}
\usepackage{amsmath}
\usepackage{mathtools}
\usepackage{filecontents}
\usepackage{lipsum,color}
\usepackage{amssymb}
\usepackage{float}

\usepackage{amsthm}

\usepackage{algorithm}        
\usepackage{algorithmic}      

\usepackage{times} 
\usepackage{amsthm}  

\usepackage{amsfonts}

\usepackage{booktabs}

\usepackage{makecell}

\theoremstyle{break}

\begin{document}
\title{All-Optical Inter-Satellite Relays with Intelligent Beam Control: Harnessing Liquid Lenses and Optical Hard Limiters}

\author{Mohammad~Taghi~Dabiri,~Mazen~Hasna,~{\it Senior Member,~IEEE},\\
	~Saud~Althunibat,~{\it Senior Member,~IEEE},~and~Khalid~Qaraqe,~{\it Senior Member,~IEEE}
	\thanks{Mohammad Taghi Dabiri and Mazen Hasna are with the Department of Electrical Engineering, Qatar University, Doha, Qatar (e-mail: m.dabiri@qu.edu.qa; hasna@qu.edu.qa).}
	\thanks{S. Althunibat is with the Department of Communication Engineering, Al-Hussein Bin Talal University, Ma’an 23874, Jordan (e-mail: saud.althunibat@ahu.edu.jo).}
	\thanks{Khalid Qaraqe is with the College of Science and Engineering, Hamad Bin Khalifa University, Doha, Qatar (E-mail: kqaraqe@hbku.edu.qa).}
	\thanks{This publication was made possible by NPRP14C-0909-210008 from the Qatar Research, Development and Innovation (QRDI) Fund (a member of  Qatar Foundation). The statements made herein are solely the responsibility of the author[s].}
}

\maketitle
\begin{abstract}

Low Earth orbit (LEO) satellite constellations are emerging as a key enabler of next-generation communications, offering global coverage and significantly lower latency compared to traditional terrestrial networks and geostationary satellites. However, further latency reduction is essential for time-critical applications such as real-time sensing, autonomous systems, and interactive services. One critical bottleneck is the optical-to-electrical (O/E) and electrical-to-optical (E/O) conversions at intermediate nodes in multi-hop links, which introduce unwanted processing delays. To address this, we investigate an all-optical relay system based on Optical Hard Limiters (OHL), which operate purely in the optical domain to suppress noise and restore signal quality without requiring O/E conversions. First, we present a rigorous analysis of inter-satellite multi-relay communication under the OHL relaying architecture, comparing it against conventional Amplify-and-Forward (AF) and Decode-and-Forward (DF) schemes. Through this comparison, we highlight both the advantages and limitations of OHL relays, including their particular sensitivity to parameter choices such as the threshold setting and divergence angle at the transmitter. Recognizing that a LEO constellation is inherently time-varying—satellites move relative to one another, causing continuous changes in link distances and tracking errors—we propose a joint optimization strategy. This scheme adaptively tunes the OHL decision threshold and beam divergence in real time to maintain optimal performance, ultimately lowering error rates and latency. Extensive simulations in a large-scale LEO network demonstrate the viability of our method and offer insights into practical implementation for next-generation inter-satellite communication systems.

\end{abstract}


\IEEEpeerreviewmaketitle

\section{Introduction}
Satellite communication networks play a pivotal role in global connectivity, supporting a wide range of applications such as remote sensing, navigation, disaster management, and broadband internet access \cite{khammassi2023precoding, mahboob2024revolutionizing}. These networks are critical for bridging the digital divide and enabling access to modern technologies in remote and underserved regions. The importance of satellite networks is further amplified with the advent of emerging technologies like 6G, which demand ultra-reliable, low-latency communication to support applications such as autonomous vehicles, remote healthcare, augmented reality, and real-time telepresence \cite{10681506}.

Latency is one of the most critical performance parameters in communication networks, as it directly impacts the quality of service for real-time applications \cite{yigit2024digi}. Future applications dependent on 6G and beyond require end-to-end latency on the order of milliseconds, which poses significant challenges for existing terrestrial and geostationary satellite networks. Both satellite networks and terrestrial fiber-optic infrastructure rely on laser communication technologies for high-speed data transmission \cite{choudhary2024inter, dabiri2024modulating}. However, there are fundamental differences in their operational characteristics that influence latency performance \cite{chen2022delay}.

Fiber-optic links, while capable of achieving high data rates, are inherently limited by the need for repeaters approximately every 50-80 kilometers due to signal attenuation \cite{elsherif2022optical}. These repeaters introduce delays caused by signal regeneration and routing at intermediate nodes. In contrast, low Earth orbit (LEO) satellite constellations employ inter-satellite links (ISLs) that span hundreds to thousands of kilometers \cite{vieira2023modulation}. These longer links reduce the need for frequent relaying, thereby minimizing routing delays and simplifying network design. Furthermore, the propagation speed of light in a vacuum—the medium for satellite laser communication—is slightly faster than in optical fibers, offering an intrinsic advantage for reducing latency.

Recognizing these advantages, major companies such as SpaceX, OneWeb, and Amazon have launched ambitious projects like Starlink, OneWeb constellation, and Project Kuiper to deploy LEO satellite networks \cite{lagunas2024low}. These systems aim to create a global infrastructure capable of providing high-speed, low-latency connectivity even in remote areas. Their adoption is expected to revolutionize communication by offering a viable alternative to traditional terrestrial networks and geostationary satellites, particularly for latency-sensitive applications.

Laser communication technology is considered the most suitable option for high-speed and high-capacity data relay in inter-satellite links due to its inherent advantages, including high bandwidth, low power consumption, and secure communication through narrow beam divergence \cite{lee2021connectivity,chaudhry2021laser,chaudhry2023laser}. Unlike traditional radio frequency (RF) links, optical links offer significantly higher data rates while reducing interference and susceptibility to eavesdropping. However, the narrow beamwidth of laser communication, while beneficial for power efficiency and security, makes it highly sensitive to tracking system errors \cite{kahraman2021investigation,dabiri2018channel}. This issue is particularly critical in LEO constellations, where satellites move rapidly relative to each other, requiring highly precise tracking mechanisms.
As the inter-satellite link distance increases, the impact of tracking errors becomes more pronounced, leading to substantial degradation in link capacity and severe fading in the received signal. To mitigate these issues, relay-based communication is essential for breaking long links into shorter segments, improving signal quality and maintaining reliable data transmission \cite{hylton2023laser,israel2018next}. 

Although optical relays have been extensively studied in the context of Free-Space Optical (FSO) communication and fiber-optic networks \cite{dabiri2018performance,liu2020relay, erdogan2021secrecy}, their investigation in inter-satellite communication remains relatively limited. In general, relay-assisted communication can be categorized into two main types: Decode-and-Forward (DF) relays and All-Optical Amplify-and-Forward (AF) relays. 

DF relays perform optical-to-electrical (O/E) conversion, process the received signal in the electrical domain, and then electrical-to-optical (E/O) conversion before retransmitting the signal. This method provides the highest performance since, at each relay node, the signal is fully recovered, processed, and retransmitted, ensuring high signal integrity \cite{dabiri2019optimal}. However, DF relays face two major challenges. First, the computational complexity and processing overhead at the relay increase significantly, particularly as data transmission rates grow \cite{dabiri2021uav}. Second, the conversion between optical and electrical domains introduces substantial latency, making DF relays less suitable for latency-sensitive applications.

The second category comprises all-optical relays, where the signal is amplified and forwarded in the optical domain using optical amplifiers such as Erbium-Doped Fiber Amplifiers (EDFA)  \cite{yang2014performance,nor2016experimental,bayaki2012edfa}. In this approach, no additional processing or optical-to-electrical conversions occur, making it highly suitable for low-latency communication scenarios. However, the major drawback of all-optical relays is noise accumulation. As optical amplifiers amplify both the signal and noise at each relay stage, the overall signal quality degrades significantly over multiple relay hops, making them unsuitable for long-distance multi-relay systems.

A promising technology that has been widely utilized in fiber-optic networks is the Optical Hard Limiter (OHL), which performs decision-making directly in the optical domain \cite{trinh2015all,vu2018all,trinh2015all}. OHL is particularly effective for intensity-modulated signals such as On-Off Keying (OOK). Unlike all-optical amplification, OHL applies a thresholding mechanism to suppress noise accumulation and restore the original signal without requiring electrical processing. The key advantage of OHL is that it eliminates the need for O/E and E/O conversions, significantly reducing processing delays and computational complexity. Moreover, by optimizing the decision threshold, OHL has been shown to achieve performance close to that of DF relays in fiber-optic networks.

However, an open question remains: Is OHL technology suitable for inter-satellite links, where link lengths and channel conditions change dynamically? 
Although OHL technology has been extensively studied in fixed-link optical systems such as fiber-optic networks and terrestrial FSO communication, its potential in mobile communication environments remains largely unexplored—let alone its applicability to highly dynamic inter-satellite links in Low Earth Orbit (LEO). Unlike static optical links, where channel conditions remain relatively stable, LEO inter-satellite networks introduce a fundamentally different challenge: continuous variations in link distance, beam divergence, and tracking accuracy due to the orbital motion of satellites. These rapid and persistent fluctuations directly impact the optimal threshold selection in OHL-based relays, making conventional designs unsuitable for such environments. In this paper, we demonstrate that applying OHL technology with traditional design principles, as used in terrestrial optical networks, leads to severe performance degradation in inter-satellite links. The dynamic nature of LEO networks not only complicates the threshold optimization process but also introduces fundamental challenges that prevent OHL-based relays from maintaining the required quality of service. This motivates the need for an adaptive, real-time optimization framework tailored specifically for mobile optical links, ensuring robust and efficient inter-satellite communication in next-generation space networks.

\subsection{Contributions of This Work}
In this paper, we provide a comprehensive study of OHL relays for low-latency, high-throughput inter-satellite communication. The major contributions can be summarized as follows:

\begin{itemize}

	\item \textit{Analytical modeling of OHL-based relaying:} We develop a unified framework that characterizes the end-to-end performance of multi-hop OHL relay networks in LEO. The analysis accounts for misalignment-induced channel fluctuations, background noise accumulation, and the effect of the OHL decision threshold on the bit-error rate. Using these considerations, we derive closed-form error probability approximations under varying orbital conditions.
	
	\item \textit{Comparison with AF and DF systems:}
	After presenting our detailed analysis of OHL-based relays, we compare their performance with conventional AF and DF systems to highlight both the benefits and the challenges. Through this comparison, we underscore the importance of optimal parameter selection—including threshold setting and beam divergence angle—in maintaining reliable, low-latency inter-satellite communication under the dynamic conditions of LEO constellations.
	
	\item \textit{Dual-parameter optimization for threshold and beam divergence:}
	We propose an iterative two-loop algorithm to jointly refine the OHL decision threshold and beam width. By minimizing noise buildup while maintaining near-optimal error performance, the system adapts effectively to changes in satellite positions and link distances.
	
	\item \textit{Integration of tunable liquid lenses for real-time beam control:}
	We introduce a method for on-the-fly beam divergence adjustment via liquid lenses, deriving closed-form expressions that map focal length to the desired beam width. By dynamically tuning the lens voltage, the system compensates for frequent fluctuations in link distance and tracking accuracy.
	
    \item \textit{Validation in large-scale LEO constellations:} We evaluate the proposed system in a multi-plane LEO network with both intra-orbit and inter-orbit links, comparing our sub-optimal approach against exhaustive-search results. The simulation outcomes confirm that the proposed optimization algorithm accurately adapts key parameters with significantly lower computational overhead, making it suitable for real-time operation in dynamic LEO constellations where the topology changes rapidly.
\end{itemize}

The remainder of the paper is organized as follows. Section~II describes the system model and key assumptions. Section~III provides the analytical framework for OHL-based relaying and comparisons with AF and DF schemes. Section~IV presents the proposed optimization algorithm. Section~V discusses implementation aspects, including the use of tunable liquid lenses. Section~VI reports simulation results, and Section~VII concludes the paper and outlines directions for future work.

\begin{figure}
	\begin{center}
		\includegraphics[width=3.4 in]{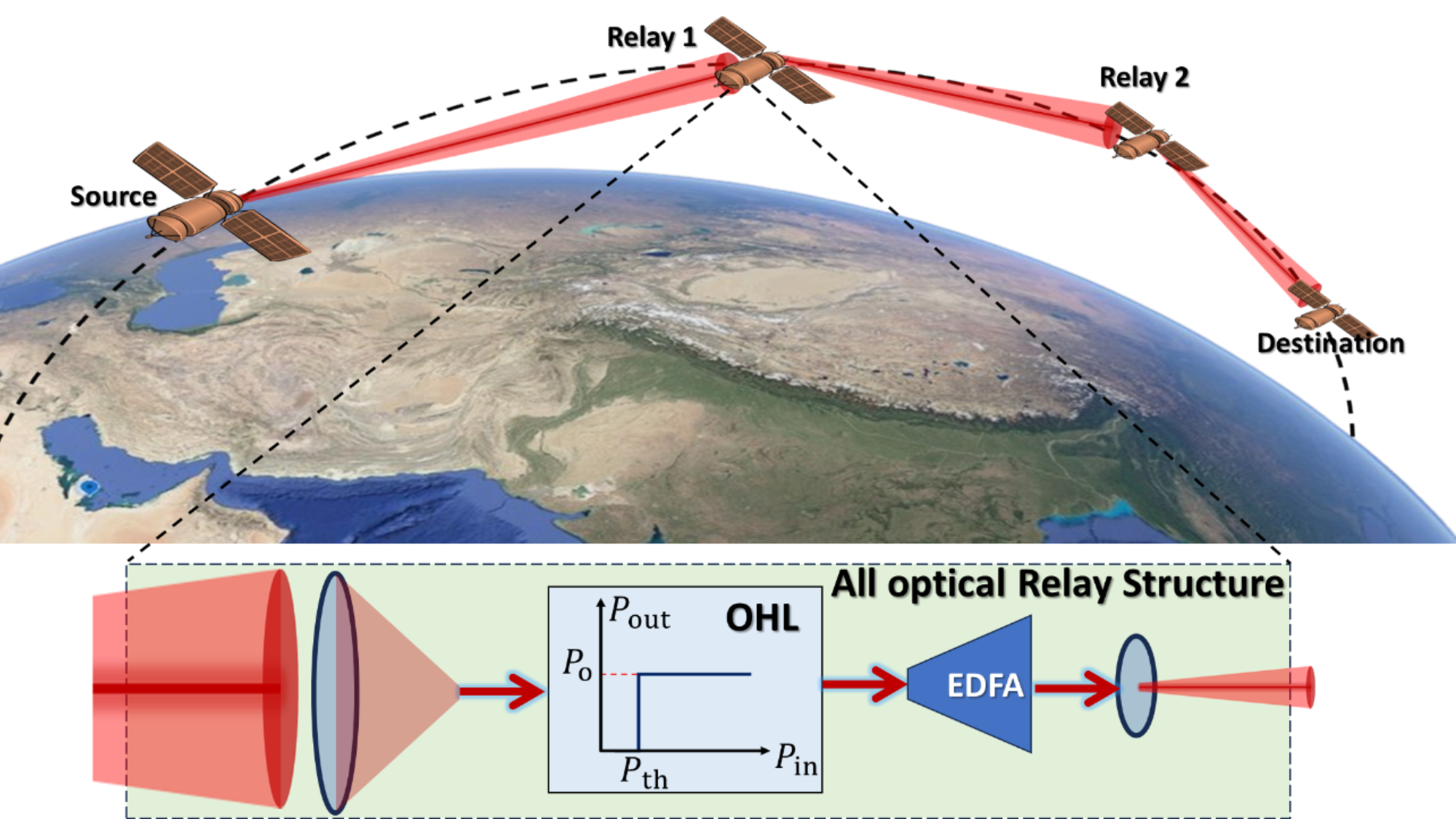}
		\caption{An illustration of a LEO satellite-based relay system for long-distance data transmission, equipped with OHL technology to eliminate optical-to-electrical and electrical-to-optical conversions, reducing signal conversion delays and minimizing the electrical processing overhead at the relays.}
		\label{se1}
	\end{center}
\end{figure}

\section{System Model}
We consider a long communication link of length $L_t$ between a source satellite and a destination satellite. This link is relayed through multiple intermediate satellites. Let $N_r$ represent the number of relays, and $R_i$ denote the $i$th relay, where $i \in \{1, \dots, N_r\}$. The communication path between the source and the destination consists of $N_r+1$ distinct links, with the length of each link denoted as $L_i$, where $i \in \{1, \dots, N_r+1\}$. 

The satellites are interconnected using high data rate laser communication technology. The primary goal is to ensure that data is transmitted from the source to the destination with minimal delay and minimal energy consumption for signal preprocessing at the relays, which is particularly important for LEO constellations. It is important to note that the problem of determining the optimal relay path is beyond the scope of this paper. We assume that such a path has already been determined at time $t$ using an external mechanism. 

In this study, the focus is on minimizing the delay caused by optical-to-electrical (O/E) and electrical-to-optical (E/O) conversions at the satellite relays. A common approach to address this issue is the use of all-optical relays based on Amplify-and-Forward (AF) techniques using Erbium-Doped Fiber Amplifiers (EDFA). An alternative method involves employing a hybrid system combining OHLs with EDFA. Both approaches are thoroughly modeled in the subsequent sections.

Since the proposed OHL-based relay system combines characteristics of both AF and DF relays, we first provide a detailed modeling of AF and DF relay systems in the following section. In the next section, we shift our focus to the analysis of the OHL-based relay system, evaluating its advantages and limitations in inter-satellite communication.


\subsection{AF Relaying Model with EDFA}
In the proposed system, the AF relaying technique is utilized with EDFAs. The system employs OOK modulation for simplicity and efficiency. The transmitted signal is represented as an optical pulse for bit ``1" and no transmission for bit ``0". 

At each relay, the received optical signal comprises the amplified data signal, background noise, and amplified spontaneous emission (ASE) noise introduced by the EDFA. The total optical power at the relay input is given as:
\begin{align}
	P_{\text{in},i} &= P_{t,i'} h_i + P_{\text{bg},i} , \nonumber \\
	P_{t,i'}& = G_i' P_{\text{in},i'} + P_{\text{ASE},i'}\label{eq:relay_input} 
\end{align}
where \( P_{t,i'}  \) represents the transmitted power from the previous node (relay \( i' = i-1 \)), $G_i$ is the EDFA gain, and \( h_i \) denoting the channel's instantaneous coefficient for the \( i \)-th link. For the first relay (\( i=1 \) or $i'=0$), \( P_{t,i'} = P_t \), where \( P_t \) is the source's transmitted power. Additionally, \( P_{\text{bg},i} \) is the background noise power modeled as additive white Gaussian noise (AWGN), and \( P_{\text{ASE},i} \) is the ASE noise power contributed by the EDFA.
The gain of the EDFA at each relay is given by:
\begin{align}
	G_i = \frac{\bar{P}_{t,i}}{P_{\text{in},i}}, \label{eq:edfa_gain}
\end{align}
where \( \bar{P}_{t,i} \) denotes the target average output power of relay \( i \), which is used to maintain a constant transmitted power level across all relays.
The ASE noise introduced by the EDFA is expressed as:
\begin{align}
	P_{\text{ASE},i} = n_{\text{sp}} h f B_0 G_i, \label{eq:ase_noise}
\end{align}
where \( n_{\text{sp}} \) is the spontaneous emission factor, \( h \) is Planck's constant, \( f \) is the optical frequency, and \( B_0 \) is the amplifier bandwidth.
The signal power received at the destination after \( N_r \) relays is obtained by recursively substituting the output of each relay into the next hop. Using \eqref{eq:relay_input}, the received power at the destination can be expressed as:
\begin{align}
	&P_{\text{dest}} = \nonumber \\ 
	&\left\{
	\begin{aligned}
		&\left( G_{N_r} P_{\text{in},N_r} + P_{\text{ASE},N_r} \right) h_{N_r+1} + P_{\text{bg},N_r+1}, \\
		&P_{\text{in},N_r} = \left( G_{N_r-1} P_{\text{in},N_r-1} + P_{\text{ASE},N_r-1} \right) h_{N_r} + P_{\text{bg},N_r}, \\
		&\hdots \\
		&P_{\text{in},2} = \left( G_1 P_{\text{in},1} + P_{\text{ASE},1} \right) h_2 + P_{\text{bg},2}, \\
		&P_{\text{in},1} = P_t h_1 + P_{\text{bg},1}.
	\end{aligned}
	\right. \label{eq:dest_signal}
\end{align}

Although the AF relay architecture operates entirely in the optical domain across intermediate nodes, the received optical signal at the final destination must be converted into an electrical signal for bit detection. This is performed by a photodetector, which produces an output current proportional to the received optical power. The resulting signal is modeled as:
\begin{align}
	i_{\text{out}} = R P_{\text{dest}} + n_{\text{th}}, \label{eq:photodetector_output}
\end{align}
where \( R \) is the responsivity of the photodetector, and \( n_{\text{th}} \sim \mathcal{N}(0, \sigma_{n,th}^2) \) denotes the thermal noise component, modeled as a zero-mean Gaussian random variable with power spectral density \( \sigma_{n,th}^2 \).

\subsection{Inter-Satellite Channel Model}
The inter-satellite laser communication channel is modeled considering a Gaussian beam profile and a circular aperture. In the vacuum of space, the dominant factors affecting the received optical power are beam spreading, alignment accuracy, and receiver aperture size.
A Gaussian beam propagates through space with an initial beam waist \( w_{0i} \) at the transmitter for the \( i \)-th link. The beam's radius at a distance \( L_i \) (the link length for the \( i \)-th segment) is given by:
\begin{align}
	w_i = w_{0i} \sqrt{1 + \left(\frac{\lambda L_i}{\pi w_{0i}^2}\right)^2}, \label{eq:beam_radius}
\end{align}
where \( \lambda \) is the wavelength of the laser. 
Given that the satellites are moving at high velocities, the length of the link \( L_i \) is constantly changing due to relative motion. This dynamic variation directly impacts the beam radius \( w_i \), as defined in \eqref{eq:beam_radius}. \footnote{To ensure efficient communication, the beam waist \( w_{0i} \) at the transmitter must be carefully controlled, as it determines the divergence and intensity of the beam over the varying link length. Optimizing \( w_{0i} \) is critical to balancing beam spreading and ensuring maximum received power at the destination.}

The inter-satellite channel gain is determined by the fraction of optical power collected by the receiver aperture relative to the total power distributed across the Gaussian beam pattern.\footnote{Since OHL systems are based on intensity modulation and direct detection (IM/DD), they are inherently insensitive to frequency or phase shifts (e.g., Doppler effects), as the photodetector responds only to instantaneous optical power \cite{hemmati2020near,ghassemlooy2017visible}.}
Additionally, tracking errors at the transmitter introduce angular deviations \( \theta_e = (\theta_x, \theta_y) \), which cause the beam center to deviate from the center of the receiver aperture. 
Let the circular receiver aperture have a radius \( r_a \). The Gaussian beam intensity at a point \( (x, y) \) in the plane of the receiver aperture is given by:
\begin{align}
	I(x, y) = \frac{2 P_t}{\pi w_i^2} \exp\left(-\frac{2 \left[(x - x_c)^2 + (y - y_c)^2\right]}{w_i^2}\right), \label{eq:beam_intensity}
\end{align}
where \( P_t \) is the transmitted power, \( w_i \) is the beam radius at the receiver (defined in \eqref{eq:beam_radius}), and \( (x_c, y_c) \) represents the beam center, which deviates from the aperture center due to the tracking errors \( \theta_e = (\theta_x, \theta_y) \). The displacement of the beam center is given by:
\begin{align}
	x_c = L_i \theta_x, \quad y_c = L_i \theta_y. \label{eq:beam_center}
\end{align}
The instantaneous channel coefficient is defined as the ratio of the received optical signal power to the transmitted optical power which is modeled as:
\begin{align}
	&h_i = \frac{2 }{\pi w_i^2} \int_{-r_a}^{r_a} \int_{-\sqrt{r_a^2 - x^2}}^{\sqrt{r_a^2 - x^2}}  \nonumber \\
	& ~~~~~~\exp\left(-\frac{2 \left[(x - L_i \theta_x)^2 + (y - L_i \theta_y)^2\right]}{w_i^2}\right) dy \, dx. \label{eq:received_power_displaced}
\end{align}

Since the tracking errors \( \theta_e = (\theta_x, \theta_y) \) are modeled as random angular deviations, the channel coefficient \( h_i \) also becomes a random variable. Typically, \( \theta_x \) and \( \theta_y \) are assumed to follow independent zero-mean Gaussian distributions with variance \( \sigma_\theta^2 \) \cite{safi2024cubesat}:
\begin{align}
	\theta_x \sim \mathcal{N}(0, \sigma_\theta^2), \quad \theta_y \sim \mathcal{N}(0, \sigma_\theta^2). \label{eq:tracking_error_distribution}
\end{align}
The random nature of \( h_i \) due to \( \theta_e \) significantly impacts the system performance, as it introduces fluctuations in the received power. Accurate modeling and mitigation of these effects are critical for reliable inter-satellite communication.

\section{Modeling and Analysis of Optical Hard Limiter}
In the context of AF relaying with EDFA, while avoiding O/E and E/O conversions simplifies the system and reduces signal processing requirements, the performance suffers due to noise accumulation. Background noise and ASE noise introduced by EDFAs in previous relays are amplified and propagate through the system. This degradation impacts the quality of the received signal at the destination, especially for long communication links with multiple relays.
To address this challenge, an OHL is proposed as a preprocessing block at each relay. The OHL is designed to mitigate the effect of noise by estimating and restricting the optical signal intensity, thus preventing excessive noise power from propagating through subsequent relays. 

The input optical signal to the OHL at relay \(i\) is represented as:
\begin{align}
	P_{\text{in},i} = P_{t,i'} h_i + P_{\text{bg},i}, \label{eq:ohl_input_signal}
\end{align}
It is assumed that the signal \(s_k \in \{0, 1\}\), modulated as OOK, passes through the OHL for optical decision-making based on intensity thresholds.\footnote{The OHL architecture is inherently designed for IM/DD, which naturally aligns with simple binary modulation formats such as OOK. This is because the optical OHL operates directly on the received optical power. More complex modulation schemes that rely on phase or frequency variation are not compatible with the OHL structure.}

The OHL performs its decision-making entirely in the optical domain, relying on the inherent properties of light intensity and thresholding to classify the incoming signal. By avoiding O/E and E/O conversions, the OHL reduces complexity and latency, while maintaining compatibility with high-speed optical systems. This optical decision-making process leverages passive optical elements, such as saturable absorbers or optical limiters, to impose thresholds on the received signal intensity. 
After the OHL decision process, the output signal \(P_{\text{OHL},i}\) is given by:
\begin{align}
	P_{\text{OHL},i} = 
	\begin{cases} 
		P_o, & P_{\text{in},i} \geq P_{\text{th}}, \\
		0, & P_{\text{in},i} < P_{\text{th}}.
	\end{cases} \label{eq:ohl_output_signal_single_threshold}
\end{align}
The OHL output, \( P_{\text{OHL},i} \), is then passed through an EDFA with fixed gain \( G \). The resulting optical signal at the EDFA output is given by:
\begin{align}
	P_{t,i} = G P_{\text{OHL},i} + P_{\text{ASE},i}, \label{eq:ohl_edfa_output}
\end{align}
where \( P_{\text{ASE},i} \) is defined in \eqref{eq:ase_noise}. Using \eqref{eq:ohl_edfa_output}, the received optical power at relay \( i \) can be rewritten as:
\begin{align}
	P_{\text{in},i} = G P_{\text{OHL},i'} h_i + P_{\text{ASE},i'} h_i + P_{\text{bg},i}. \label{eq:ohl_input_signal_2}
\end{align} 
In \eqref{eq:ohl_input_signal_2}, the ASE noise term \( P_{\text{ASE},i'} \) from the previous EDFA is attenuated by the channel coefficient \( h_i \), and can therefore be well approximated as:
\begin{align}
	P_{\text{in},i} \simeq G P_{\text{OHL},i'} h_i + P_{\text{bg},i}. \label{eq:ohl_input_signal_3}
\end{align} 
Such an approximation does not hold in conventional AF relaying, where the accumulated ASE noise \( P_{\text{ASE},i'} h_i \) is passed directly into the next EDFA stage and further amplified. In contrast, in the OHL-based system, this term is passed through the OHL, where a hard decision is made based on total input power, as defined in \eqref{eq:ohl_output_signal_single_threshold}. Thus, noise accumulation is effectively suppressed before amplification.

The OHL's ability to act directly in the optical domain eliminates unnecessary conversions and enables real-time intensity-based decision-making. This feature significantly reduces noise accumulation in the system by ensuring that only signals meeting the threshold criteria propagate to the next stage of the relay chain.
However, the decision-making process in the OHL is inherently hard, as it relies on a fixed threshold \(P_{\text{th}}\) to classify the incoming signal. Unlike DF relays, where the threshold or decision boundaries can dynamically adapt based on channel state information (CSI), the fixed-threshold nature of the OHL makes it less robust to severe channel fluctuations.

\subsection{Error Probability}
The decision-making process in the OHL introduces a possibility of misclassification due to noise and channel variations. For a single relay link \(i\), the probability of error depends on the received optical power \(P_{\text{in},i}\), the fixed threshold \(P_{\text{th}}\), and the random channel coefficient \(h_i\). The error probability is expressed as:
\begin{align}
	\mathbb{P}_{\text{e},i} &= \frac{1}{2} \left[\mathbb{P}_{\text{e},i}(s_k = 1) + \mathbb{P}_{\text{e},i}(s_k = 0)\right], \label{eq:average_error_general}
\end{align}
where 
\begin{align}
	\mathbb{P}_{\text{e},i}(s_k = 1) &= \int_0^\infty Q\left(\frac{P_{t,i'} h - P_{\text{th}}}{\sigma_{\text{bg},i}}\right) f_{h_i}(h) \, dh,  \\
	\mathbb{P}_{\text{e},i}(s_k = 0) &= \int_0^\infty  Q\left(\frac{P_{\text{th}}}{\sigma_{\text{bg},i}}\right) f_{h_i}(h) \, dh, \label{sx1}
\end{align}
where \( Q(x) \) is the Q-function.

In practice, for inter-satellite communication links, the Gaussian beam radius \(w_i\) at the receiver plane is typically in the order of tens of meters, which is significantly larger than the radius of the receiver aperture \(r_a\). As a result, the variation of the optical intensity over the surface of the receiver aperture is negligible. 
Thus, \eqref{eq:received_power_displaced} can be greatly simplified as:
\begin{align}
	h_i = \frac{r_a^2}{w_i^2} \cdot \exp\left(-\frac{2 \left(L_i^2 \theta_x^2 + L_i^2 \theta_y^2\right)}{w_i^2}\right). \label{eq:simplified_channel_gain}
\end{align}
%
Based on \eqref{eq:tracking_error_distribution}, \( z = \theta_x^2 + \theta_y^2 \) follows a chi-squared distribution with 2 degrees of freedom, equivalent to an exponential distribution:
\begin{align}
	f_Z(z) = \frac{1}{2 \sigma_\theta^2} \exp\left(-\frac{z}{2 \sigma_\theta^2}\right), \quad z \geq 0.
\end{align}
By substituting \( z = \theta_x^2 + \theta_y^2 \) into \( h_i \), we find:
\begin{align}
	z = -\frac{w_i^2}{2L_i^2} \ln\left(\frac{h_i w_i^2}{r_a^2}\right).
\end{align}
The CDF of \( h_i \) is defined as:
\begin{align}
	F_{h_i}(h_i) &= P(H_i \leq h_i) = P\left(Z \geq -\frac{w_i^2}{2L_i^2} \ln\left(\frac{h_i w_i^2}{r_a^2}\right)\right) \nonumber \\
	&= \exp\left(\frac{w_i^2}{4L_i^2 \sigma_\theta^2} \ln\left(\frac{h_i w_i^2}{r_a^2}\right)\right).
\end{align}
The PDF is derived by differentiating the CDF with respect to \( h_i \):
\begin{align}
	f_{h_i}(h_i)& = \frac{d}{d h_i} F_{h_i}(h_i) \nonumber \\
	&= \frac{w_i^2}{4 L_i^2 \sigma_\theta^2 h_i} \left(\frac{h_i w_i^2}{r_a^2}\right)^{\frac{w_i^2}{4 L_i^2 \sigma_\theta^2}}, \quad 0<h_i < \frac{r_a^2}{w_i^2}.
	\label{eq:pdf_channel_gain}
\end{align}
Substituting \eqref{eq:pdf_channel_gain} into \eqref{sx1}, the error probability of the desired system is obtained as
\begin{align}   \label{eq:pe_0_general} 
	&\mathbb{P}_{\text{e},i,OHL} = \frac{1}{2}Q\left(\frac{P_{\text{th}}}{\sigma_{\text{bg},i}}\right)   \\
	&~~~ +  \frac{\alpha }{2 } w_i^2 \left(\frac{w_i^2}{r_a^2}\right)^{\alpha     w_i^2}
	\int_0^{\frac{r_a^2}{w_i^2}}  h^{\alpha w_i^2-1} Q\left(\frac{P_{t,i'} h - P_{\text{th}}}{\sigma_{\text{bg},i}}\right)  \, dh, \nonumber
\end{align}
where \(\alpha = \frac{1}{4 L_i^2 \sigma_\theta^2}\).

The main difference between relaying using OHL and DF relaying lies in the hard and soft thresholds of these two methods. Specifically, OHL employs a hard decision mechanism. However, in DF relaying, the decision-making is performed in the electronic domain, where the channel gain can be easily estimated, and the optimized threshold can be set as \( P_{\text{th,s}} = \frac{P_{t,i} h_i}{2} \). In this case, the relationship in \eqref{eq:pe_0_general} for DF relaying is modified as follows:
\begin{align}
	\mathbb{P}_{\text{e},i,DF} &= \alpha w_i^2 \left(\frac{w_i^2}{r_a^2}\right)^{\alpha w_i^2}    
	\int_0^{\frac{r_a^2}{w_i^2}}  h^{\alpha w_i^2-1} Q\left(\frac{P_{t,i'} h }{2\sigma'_{\text{bg},i}}\right)  \, dh. \nonumber\\\label{eq:pe_0_DF}
\end{align}
where $ \sigma'_{\text{bg},i} = \sqrt{\sigma_{n,th}^2 + \sigma_{\text{bg},i}^2 }$.

The Q-function can be approximated as:
\begin{align} 
	Q(x) &\approx \sum_{j=1}^{3} a_j \exp(-b_j x^2), \label{eq:Q_approx}
\end{align}
where the coefficients are:
\begin{align}
	\left\{
	\begin{array}{ll}
		a_1 = \frac{5}{24}, & a_2 = \frac{4}{24}, \quad a_3 = \frac{1}{24},  \\ 
		b_1 = 2, & b_2 = \frac{11}{20}, \quad b_3 = \frac{1}{2}.
	\end{array}
	\right.
\end{align}
Substituting \eqref{eq:Q_approx} into \eqref{eq:pe_0_DF}, we get:
\begin{align} \label{eq:integral_exp}
	\mathbb{P}_{\text{e},i,DF} &= \alpha \left(\frac{w_i^2}{r_a^2}\right)^\alpha \sum_{j=1}^3 a_j \nonumber \\
	&\times \underbrace{\int_0^{\frac{r_a^2}{w_i^2}} h^{\alpha-1} \exp\left(-b_j \frac{P_{t,i'}^2 h^2}{4 (\sigma'_{\text{bg},i})^2}\right) dh}_{I_j}. 
\end{align}
Using the substitution $k_j = \frac{b_j P_{t,i'}^2}{4 \sigma_{\text{bg},i}^2}$, \( u = k_j h^2 \), \( du = 2 k_j h \, dh \), the integral $I_j$ becomes:
\begin{align}
	I_j &= \frac{1}{2 k_j^{\alpha/2}} \int_0^{k_j \left(\frac{r_a^2}{w_i^2}\right)^2} u^{\frac{\alpha}{2} - 1} \exp(-u) du.
\end{align}
This is the definition of the lower incomplete Gamma function:
\begin{align}
	\gamma(s, x) = \int_0^x t^{s-1} e^{-t} dt.
\end{align}
Thus:
\begin{align}
	I_j = \frac{1}{2 k_j^{\alpha/2}} \gamma\left(\frac{\alpha}{2}, k_j \left(\frac{r_a^2}{w_i^2}\right)^2\right).
\end{align}
Substituting \( I_j \) into \eqref{eq:integral_exp}, the closed-form expression for \( P_{\text{e},i,DF} \) is:
\begin{align}
	\mathbb{P}_{\text{e},i,DF} &= \alpha \left(\frac{w_i^2}{r_a^2}\right)^\alpha 
	\sum_{j=1}^3 \frac{a_j}{2 k_j^{\alpha/2}} \gamma\left(\frac{\alpha}{2}, k_j \left(\frac{r_a^2}{w_i^2}\right)^2\right), \nonumber
\end{align} 
where $k_j = \frac{b_j P_{t,i'}^2}{4 (\sigma'_{\text{bg},i})^2}$.

Note that for OHL relaying, the decision-making at the destination is also performed using a soft decision mechanism. In this case, the end-to-end error probability of a multi-relay system based on OHL is modeled as follows:
\begin{align} \label{ohl1}
	\mathbb{P}_{\text{e},E2E} &= 1 - (1-P_{\text{e},N_r+1,DF})\prod_{i=1}^{N_r}(1-P_{\text{e},i,OHL}).
\end{align}
\eqref{ohl1} is modified for DF relaying as follows:
\begin{align} \label{ohl2}
	\mathbb{P}_{\text{e},E2E} &= 1 - \prod_{i=1}^{N_r+1}(1-P_{\text{e},i,DF}).
\end{align}

\begin{table}[ht]
	\centering
	\caption{Values of simulation parameters}
	\begin{tabular}{|c c|c c|}
		\hline
		\textbf{Parameter} & \textbf{Value} & \textbf{Parameter} & \textbf{Value} \\
		\hline
		$P_{t,i}$ & 4~W & $N_r$ & 1 to 14 \\
		$\sigma_\theta$ (intra-orbit) & 50~$\mu$rad & $\sigma_\theta$ (inter-orbit) & 150~$\mu$rad \\
		$r_a$ & 10~cm & $\theta_{\text{div}}$ & 400~$\mu$rad \\
		Orbit altitude & 600~km & Inclination angle & 53$^\circ$ \\
		$h$ & $6.6 \times 10^{-34}$~J$\cdot$s & $\lambda$ & 1550~nm \\
		$f$ & $1.9 \times 10^{14}$~Hz & $B_0$ & $2 \times 10^8$~Hz \\
		$n_{\text{sp}}$ & 1.1 & $P_{\text{bg}}$ & $6 \times 10^{-9}$~W \\
		$R$ & 0.8~A/W & $\sigma_{n,\text{th}}$ & $1 \times 10^{-9}$ \\
		$L_i$ & 200--2000~km & $\sigma_\theta$ & 80--160~$\mu$rad \\
		$w_i$ & 200--600~m & $P_{\text{th}}$ & 1--100~nW \\
		$F$ & 15--60~mm & $w_0$ & 1--5~mm \\
		$z_R$ & $\frac{\pi w_0^2}{\lambda}$ & $L'$ & 30--50~mm \\
		\hline
	\end{tabular} \label{tab:sim-params}
\end{table}

\subsection{Simulation Results and Performance Analysis}

\begin{figure}
	\begin{center}
		\includegraphics[width=3.0 in]{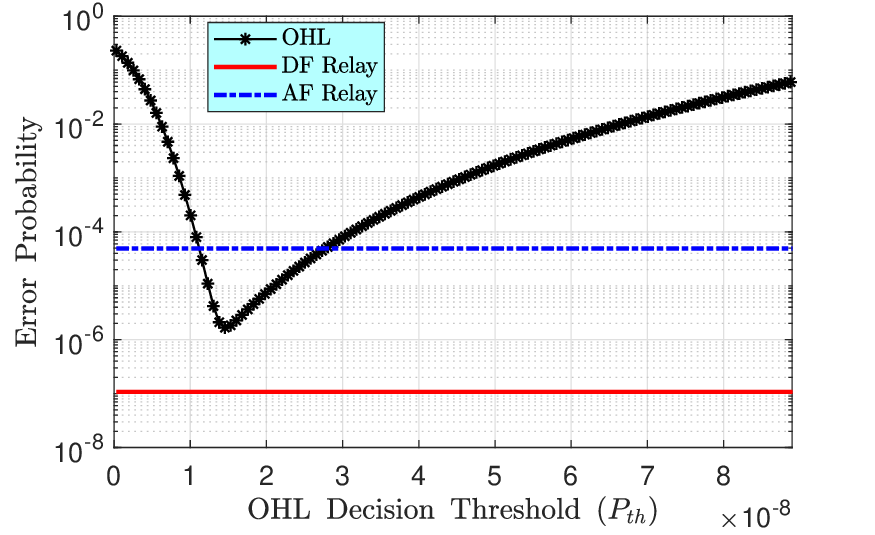}
		\caption{The error probability of a single-relay OHL system as a function of the optical hard threshold \( P_{\text{th}} \). The performance is compared with DF and AF relays.}	
		\label{sb1}
	\end{center}
\end{figure}
%

Here, we present several simulations to evaluate the impact of the relay technology employed on the performance of inter-satellite links. For the simulations, commonly used parameters from the inter-satellite communication literature are adopted \cite{kaushal2016optical}. The simulation parameters used in this paper are summarized in Table~\ref{tab:sim-params}. 

\begin{figure}
	\begin{center}
		\includegraphics[width=3.0 in]{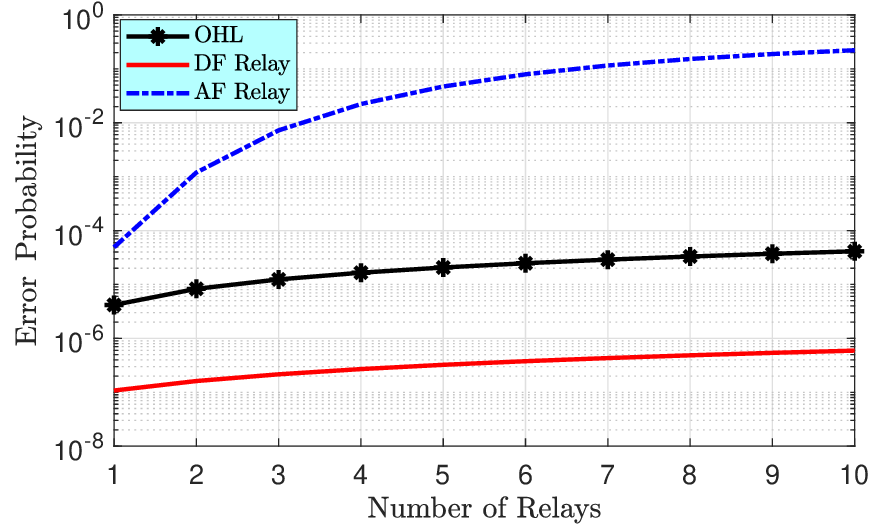}
		\caption{The error probability of multi-relay OHL, DF, and AF systems as a function of the number of intermediate relays. The OHL system uses the optimal threshold derived from Fig. \ref{sb1}. }
		
		\label{sb2}
	\end{center}
\end{figure}
%

In Fig. \ref{sb1}, the impact of the optical hard threshold \( P_{\text{th}} \) on the performance of OHL relays is analyzed. The figure illustrates the error probability of a single-relay OHL system as a function of \( P_{\text{th}} \), and its performance is compared with DF and AF relays. As expected, the DF relay achieves the best performance due to its ability to utilize the O/E and E/O conversions and advanced electronic processing, albeit at the cost of increased delay and computational complexity. On the other hand, the performance of the OHL relay is highly dependent on the value of \( P_{\text{th}} \). It is evident that selecting an optimal value for \( P_{\text{th}} \) can significantly enhance the performance of the OHL relay, even surpassing that of the AF relay under certain conditions.

\begin{figure}[t]
	\centering
	\subfloat[]{\includegraphics[width=1.65in]{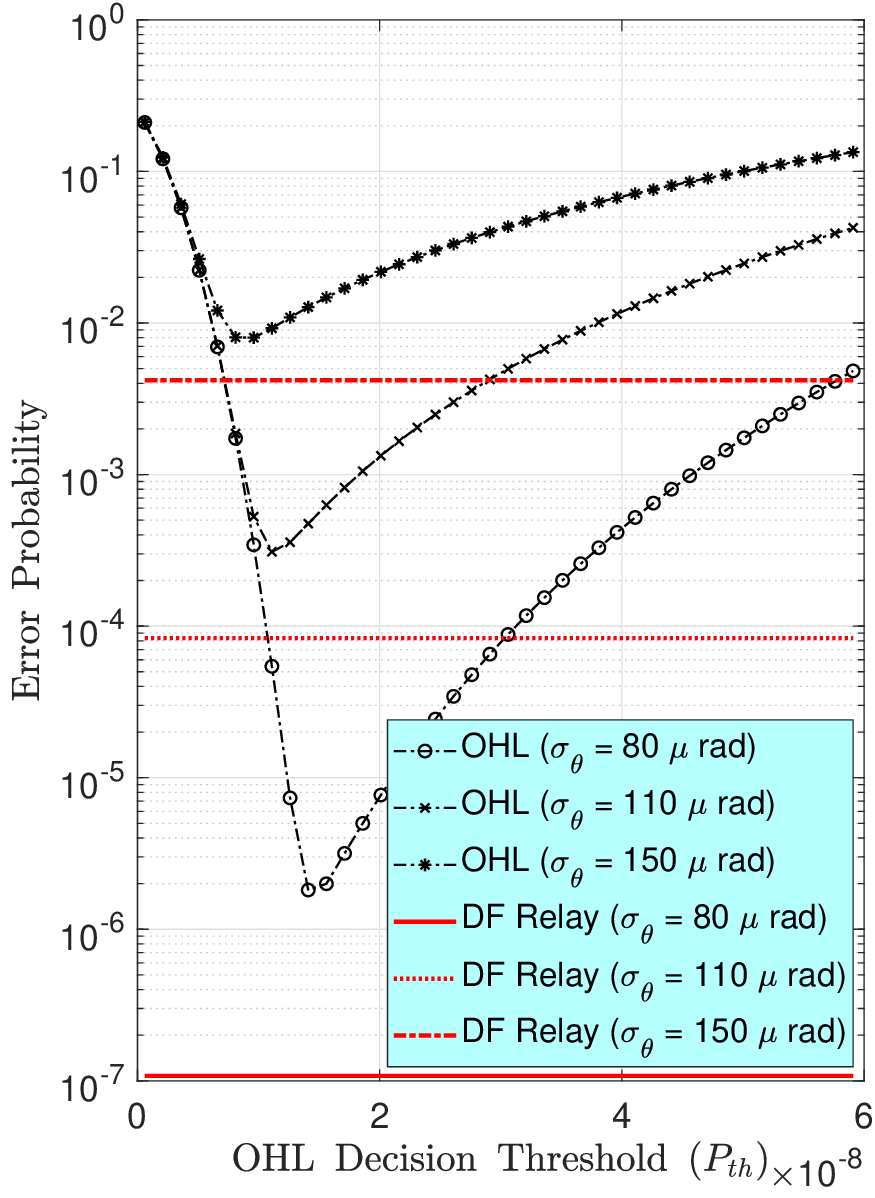}%
		\label{sb3}}
	\hfil
	\subfloat[]{\includegraphics[width=1.65in]{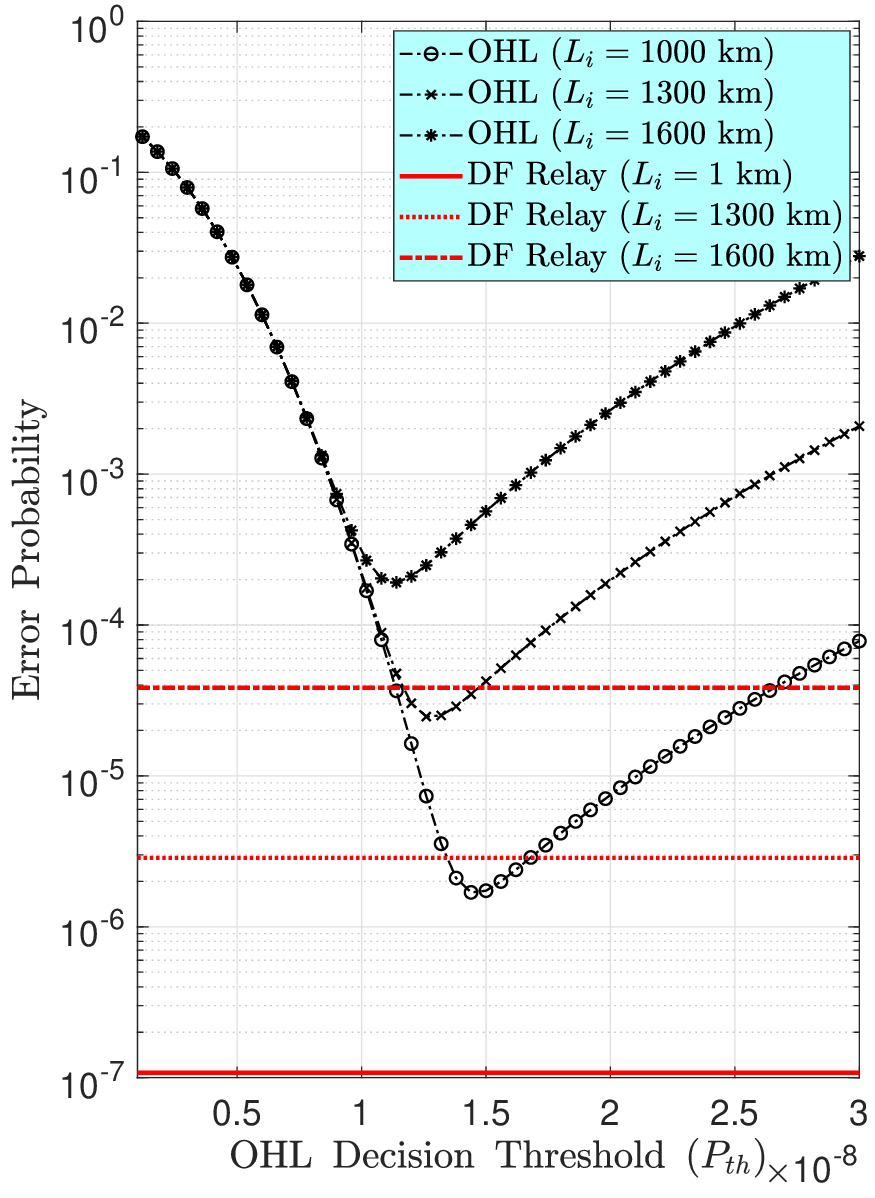}%
		\label{sn3}}
	\caption{The error probability of the OHL system as a function of the hard threshold \( P_{\text{th}} \). (a) For different tracking system accuracies (\( \sigma_\theta = 80, 110, \text{and } 160 \, \mu\text{rad} \)). (b) For different link lengths (\( L_i = 1000, 1300, \text{and } 1600 \, \text{km} \)).}
	\label{fig:combined}
\end{figure}

To provide a better comparison with the AF relay, Fig. \ref{sb2} evaluates the performance of the OHL-based relay system for a multi-relay scenario. For the OHL system, the optimal threshold value obtained from Fig. \ref{sb1} is used. 
The figure plots the error probability as a function of the number of intermediate relays between the source and destination satellites. As expected, the performance of all three relay systems improves as the number of relays decreases. 
However, the performance improvement in the AF relay decreases more significantly compared to the other relays due to noise amplification in consecutive relays. In contrast, the OHL relay, by performing an optical hard decision on the received signal intensity at each stage, achieves optical detection on the OOK-modulated bits, demonstrating greater resilience against an increasing number of relays.

It should be noted that selecting the optimal threshold value for OHL relays is a major challenge in inter-satellite communications, especially for inter-orbital satellite links where satellites move past each other at higher speeds, causing variations in link length and tracking system accuracy. To investigate this, Figs. \ref{sb3} and \ref{sn3} analyze the performance of the OHL system under different conditions as a function of the hard threshold.
The results in Fig. \ref{sb3} are plotted for three different tracking error variances, \( \sigma_\theta = 80 \, \mu\text{rad} \), \( \sigma_\theta = 110 \, \mu\text{rad} \), and \( \sigma_\theta = 160 \, \mu\text{rad} \). Meanwhile, the results in Fig. \ref{sb4} are plotted for three different link lengths, \( L_i = 500 \, \text{km} \), \( L_i = 1000 \, \text{km} \), and \( L_i = 1600 \, \text{km} \).
As shown in Fig. \ref{sb3}, increasing the tracking system's accuracy results in a higher optimal decision threshold. This is due to the reduced fluctuation in the channel gain caused by lower tracking error, allowing the decision threshold to increase and thus reducing the error caused by background noise. Similarly, as shown in Fig. \ref{sn3}, a similar trend is observed for shorter link lengths, where the optimal threshold increases as the link length decreases. 

The simulation results highlight the critical importance of selecting the optimal threshold in the proposed OHL-based relay system, particularly in dynamic inter-satellite networks where both link lengths and network topology continuously change. Unlike static terrestrial systems, inter-satellite links are subject to constant variations due to satellite motion, leading to fluctuations in beam width as a function of link distance. Consequently, continuous adjustments to the beam divergence angle are required to maintain optimal transmission efficiency.
Moreover, since the optimal beam width is inherently dependent on the optimal threshold value, any adaptation in one parameter directly influences the other. This interdependency underscores the necessity for a joint optimization approach that dynamically updates both parameters in real time.
Given these challenges, achieving an optimal and continuously adaptive design for the OHL-based relay system is crucial for ensuring reliable communication with minimal error probability. In the following sections, we focus on developing an optimization framework that effectively integrates threshold selection and beam divergence control, enabling seamless adaptation to the evolving satellite network topology and link conditions.

\section{Optimal Design of OHL-Based Systems}

In the previous section, we demonstrated that the performance of the OHL-based relay system is a function of the decision threshold \( P_{\text{th}} \) and the beam width at the receiver, which is controlled by the divergence angle at the transmitter. The selection of these parameters significantly impacts the reliability and efficiency of inter-satellite optical links. 

In this section, we focus on the optimal design of the proposed OHL-based multi-relay system for establishing long-distance inter-satellite communication links. 
In the optimal design of OHL-based relay systems, key parameters are highly interdependent. Any change in the optimal value of one parameter directly affects the optimal value of the other. This mutual dependency makes the optimization process complex and time-dependent, especially in inter-satellite communication, where link distance and channel conditions continuously change. 

Ideally, a joint optimization approach could be formulated to determine the optimal values of both \( w_i^* \) and \( P_{\text{th}}^* \) simultaneously. However, implementing a fully optimal algorithm requires extensive computational resources and real-time access to complete channel state information across all relay nodes, which is practically challenging in dynamic satellite networks. 

To balance performance and computational complexity, this section introduces a sub-optimal method for determining the optimal parameter values. The proposed approach is designed to rapidly and accurately adjust system parameters in response to real-time variations in the relay system.

\subsection{Optimal Threshold $P_{th}$}
Here, we assume that the optimal value of \( w_i^* \) is known, and we use it to determine the optimal decision threshold \( P_{\text{th}}^* \).
To find the optimal value of $P_{th}$ that minimizes the error probability $P_{e,i,OHL}$, we differentiate it with respect to $P_{th}$ and set the derivative equal to zero:

\begin{align} \label{gr0}
	& \frac{d P_{e,i,OHL}}{d P_{th}} = \frac{1}{2} \frac{d}{d P_{th}} Q\left( \frac{P_{th}}{\sigma_{bg,i}} \right) +  \\
	& \alpha \frac{2}{w_i^2} \left(\frac{w_i^2}{r_a^2} \right)^{\alpha w_i^2} 
	\frac{d}{d P_{th}}
	\int_0^{r_a^2 w_i^2} h^{\alpha w_i^2 - 1} Q\left(\frac{P_{t,i}' h - P_{th}}{\sigma_{bg,i}} \right) dh. \nonumber
\end{align}
Differentiating the two terms inside the integral:
\begin{align} \label{gr1}
	&\frac{d}{d P_{th}} \int_0^{r_a^2 w_i^2} h^{\alpha w_i^2 - 1} Q\left( \frac{P_{t,i}' h - P_{th}}{\sigma_{bg,i}} \right) dh  \nonumber \\ 
	&~~~~~= - \int_0^{r_a^2 w_i^2} h^{\alpha w_i^2 - 1} \frac{1}{\sqrt{2\pi} \sigma_{bg,i}} e^{- \frac{(P_{t,i}' h - P_{th})^2}{2 \sigma_{bg,i}^2}} dh,
\end{align}
and
\begin{equation} \label{gr2}
	\frac{d}{d P_{th}} Q\left( \frac{P_{th}}{\sigma_{bg,i}} \right) = \frac{1}{\sqrt{2\pi} \sigma_{bg,i}} e^{- \frac{P_{th}^2}{2 \sigma_{bg,i}^2}}.
\end{equation}
Applying \eqref{gr1} and \eqref{gr2} to \eqref{gr0} and setting the Derivative to Zero, we obtain:
\begin{align} \label{gr4}
	&- \alpha \frac{2}{w_i^2} \left(\frac{w_i^2}{r_a^2} \right)^{\alpha w_i^2} \int_0^{r_a^2 w_i^2} h^{\alpha w_i^2 - 1} \frac{1}{\sqrt{2\pi} \sigma_{bg,i}} e^{- \frac{(P_{t,i}' h - P_{th})^2}{2 \sigma_{bg,i}^2}} dh \nonumber \\
	&+ \frac{1}{\sqrt{2\pi} \sigma_{bg,i}} e^{- \frac{P_{th}^2}{2 \sigma_{bg,i}^2}} = 0.
\end{align}
Taking the natural logarithm on both sides of \eqref{gr4}, we obtain:
\begin{align} \label{gr6}
	&-\frac{P_{th}^2}{2 \sigma_{bg,i}^2} =  \\
	& \ln \left( \alpha \frac{2}{w_i^2} \left(\frac{w_i^2}{r_a^2} \right)^{\alpha w_i^2} \int_0^{r_a^2 w_i^2} h^{\alpha w_i^2 - 1} e^{- \frac{(P_{t,i}' h - P_{th})^2}{2 \sigma_{bg,i}^2}} dh \right). \nonumber
\end{align}
Finally, using \eqref{gr6}, the optimal threshold $P_{th}^*$ is derived as:
\begin{align} \label{gr7}
	&P_{th}^* = -2 \sigma_{bg,i}^2 \ln \Bigg( \alpha \frac{2}{w_i^2} \left(\frac{w_i^2}{r_a^2} \right)^{\alpha w_i^2}  \nonumber  \\
	&~~~~~~~~\times \int_0^{r_a^2 w_i^2} h^{\alpha w_i^2 - 1} e^{- \frac{(P_{t,i}' h - P_{th})^2}{2 \sigma_{bg,i}^2}} dh \Bigg).
\end{align}

\begin{figure}
	\centering
	\subfloat[] {\includegraphics[width=1.65 in]{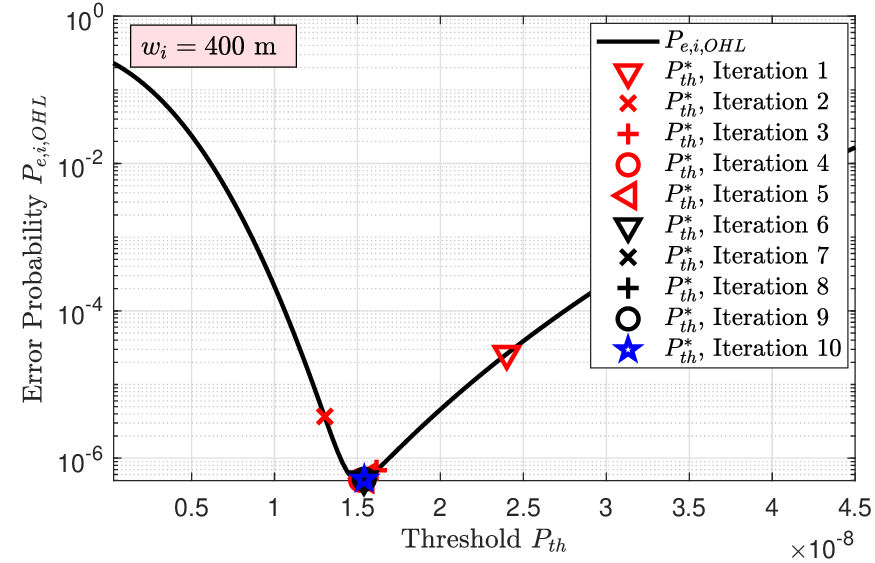}
		\label{bv1}
	}
	\hfill
	\subfloat[] {\includegraphics[width=1.65 in]{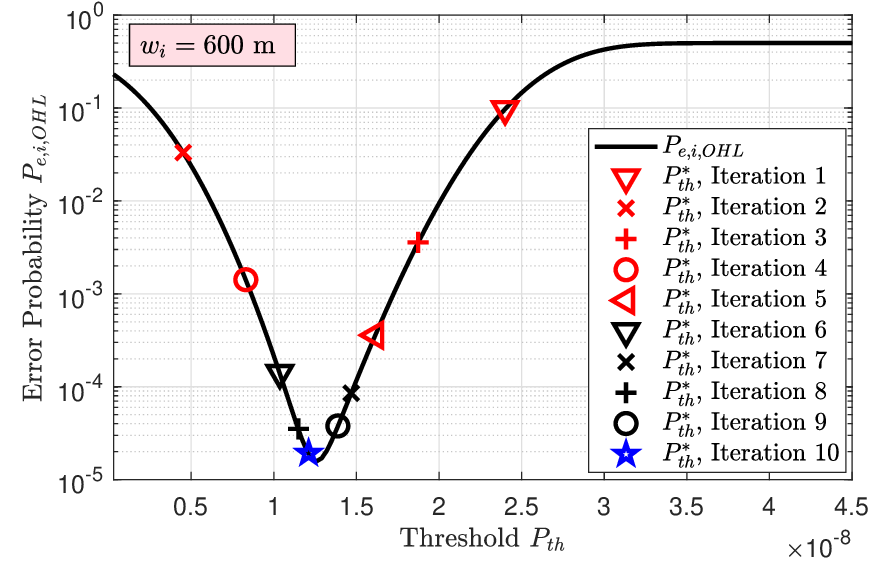}
		\label{bv2}
	}
	\caption{Convergence of the optimal threshold \( P_{\text{th}}^* \) for different values of \( w_i \). (a) \( w_i = 400 \) m and (b) \( w_i = 600 \) m. The results demonstrate that the iterative method quickly converges to the optimal threshold, with faster convergence for smaller \( w_i \) values.}
	\label{bv3}
\end{figure}

Note that the optimal threshold \( P_{\text{th}}^* \) is derived under the assumption that the optimal \( w_i^* \) is known. Moreover, \( P_{\text{th}} \) itself is a function of the threshold value \( P_{\text{th}} \). However, this relationship can be utilized iteratively: given an initial estimate of \( P_{\text{th}} \) and \( w_i \), the derived expression provides a more accurate value of \( P_{\text{th}} \), bringing it closer to the optimal value. This refined \( P_{\text{th}} \) can then be used as an updated initial value in a successive iteration, leading to a convergence towards the final optimal threshold \( P_{\text{th}}^* \) for a given \( w_i \).

To provide further insight, Fig. \ref{bv1} and Fig. \ref{bv2} illustrate the convergence behavior of \( P_{\text{th}}^* \) for two different values of \( w_i = 400 \) m and \( w_i = 600 \) m. As observed, the proposed iterative method rapidly converges to the optimal threshold \( P_{\text{th}}^* \) for a given \( w_i \). Furthermore, the convergence speed increases as \( w_i \) decreases, indicating a stronger dependence on initial conditions for smaller beam widths.
More importantly, these results demonstrate that the optimal threshold \( P_{\text{th}}^* \) is inherently a function of \( w_i \), which will be analyzed in greater detail in the following section.

\subsection{Optimization of \( w_i \)}
Given all real-time channel parameters, such as the link length, and assuming the optimal decision threshold \( P_{\text{th}}^* \) is known, the optimal beam width \( w_i^* \) can be obtained by differentiating \eqref{ohl1} and \eqref{eq:pe_0_general}. However, since \eqref{eq:pe_0_general} is a highly complex function of \( w_i \), directly solving for the optimal \( w_i^* \) from \eqref{ohl1} becomes extremely challenging.

\begin{figure}
	\centering
	\subfloat[] {\includegraphics[width=3.3 in]{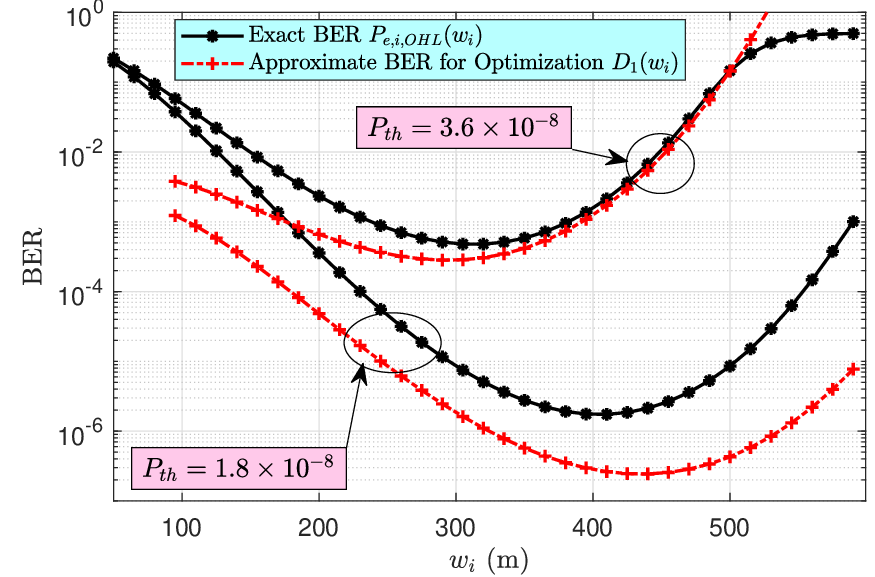}
		\label{f1}
	}
	\hfill
	\subfloat[] {\includegraphics[width=3.3 in]{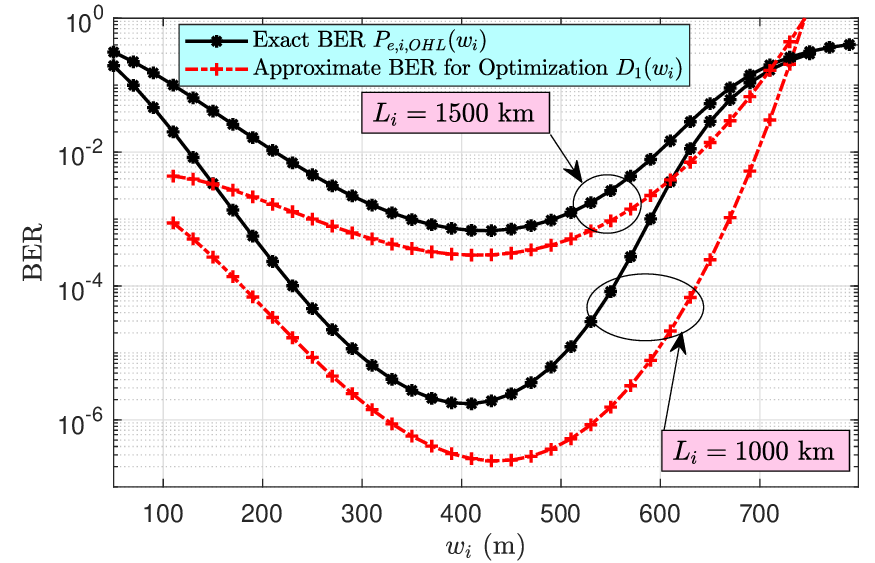}
		\label{f2}
	}
	\caption{Comparison of the exact error probability derived from \eqref{eq:pe_0_general} and the approximate model in \eqref{sd1} over a wide range of \( w_i \) values for (a) two different values of \( P_{\text{th}} \) and (b) two different link lengths. The approximation effectively follows the trend of the exact model and provides an optimal \( w_i \) that closely aligns with the true optimal value.
	}
	\label{f3}
\end{figure}

To address this issue, we approximate the behavior of \eqref{eq:pe_0_general} using a series of simplifications, leading to the following expression:
\begin{align} \label{sd1}
	P_{\text{e,aprox}} &= \left( \frac{P_{\text{th}}}{P_{t,i'}} \frac{w_i^2}{r_a^2} \right)^{\alpha w_i^2+1}.
\end{align}
To validate the accuracy of the proposed approximation, we compare the exact error probability obtained from \eqref{eq:pe_0_general} with the results derived using the approximation in \eqref{sd1} across a wide range of channel parameters, as shown in Fig. \ref{f3}.

Fig. \ref{f1} illustrates the error probability for two different values of \( P_{\text{th}} \), while Fig. \ref{f2} presents results for two different link lengths. Additionally, the figures depict the results over an extensive range of \( w_i \) values to evaluate the robustness of the approximation. 
Although the approximation in \eqref{sd1} does not perfectly match the exact values, it effectively captures the system's behavior with respect to variations in \( w_i \). Moreover, the optimal value of \( w_i \) derived from the approximation closely aligns with the actual optimal \( w_i \), demonstrating its suitability for efficient parameter tuning in real-time relay adaptation. For simplicity, let us define \( W = w_i^2 \), allowing us to rewrite \eqref{sd1} in the following form:
\begin{align} \label{sd2}
	P_{\text{e,aprox}} &= \left( \frac{P_{\text{th}}}{P_{t,i'}} \frac{W}{r_a^2} \right)^{\alpha W+1} 
	= e^{(\alpha W + 1) \ln \left( \frac{P_{\text{th}}}{P_{t,i'}} \frac{W}{r_a^2} \right)}.
\end{align}
By differentiating \eqref{sd2} and rearranging the terms, we obtain:
\begin{align} \label{sd3}
	\frac{dP_{\text{e,aprox}}}{dW} = P_{\text{e,aprox}} \left[ \alpha \ln \left( \frac{P_{\text{th}}}{P_{t,i'}} \frac{W}{r_a^2} \right) 
	 + \frac{\alpha W + 1}{W} \right] = 0.
\end{align}
\eqref{sd3} can be rewritten as follows:
\begin{align}
	W e^{\frac{1}{\alpha W}} = \frac{r_a^2}{P_{\text{th}} / P_{t,i'}} e^{-1}.
\end{align}
Using the Lambert function property, we obtain:
\begin{align} \label{sd4}
	W^* = \frac{r_a^2}{P_{\text{th}} / P_{t,i'}} e^{\text{LambertW} \left(-
		\frac{ 4 e P_{\text{th}}  L_i^2 \sigma_\theta^2 }
		{ r_a^2 P_{t,i'}} \right) - 1},
\end{align}
where $\text{LambertW}(y)$ is the Lambert function, which is supported by mathematical software such as MATLAB and Mathematica.
Using \eqref{sd4}, the optimal beam width is given by:
\begin{align} \label{sd5}
	w_i^* = \frac{r_a}{\sqrt{P_{\text{th}} / P_{t,i'}}} e^{  \frac{\text{LambertW} \left(-
			\frac{4 e P_{\text{th}}  L_i^2 \sigma_\theta^2 }
			{ r_a^2 P_{t,i'}} \right) - 1}{2} } .
\end{align}
Note that the optimal value \( w_i^* \) obtained from \eqref{sd5} is derived under the assumption that the optimal \( P_{\text{th}}^* \) is known. However, any change in \( P_{\text{th}} \) can affect the optimal value of \( w_i^* \), and vice versa.

\begin{figure}
	\begin{center}
		\includegraphics[width=3.3 in]{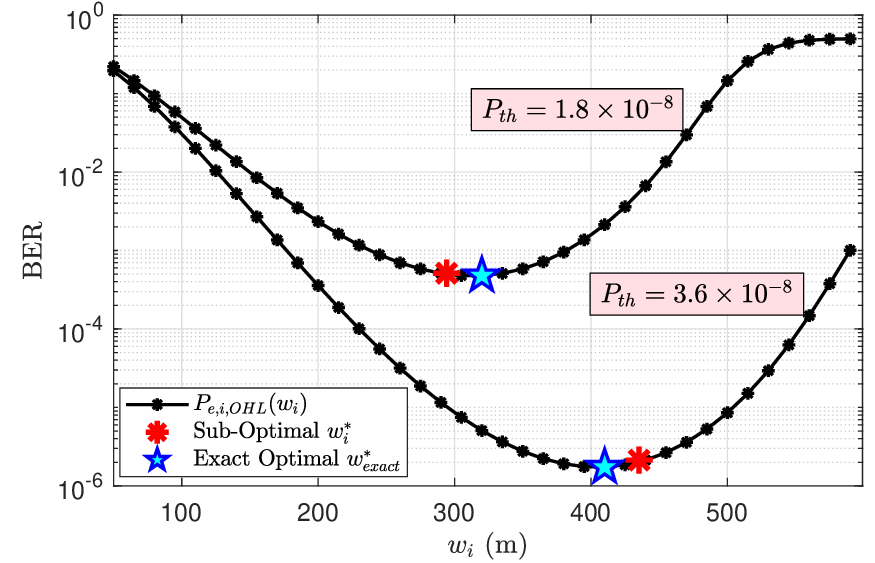}
		\caption{Comparison of the sub-optimal \( w_i \) obtained from \eqref{sd5} with the exact numerical solution for two different values of \( P_{\text{th}} \). The results illustrate the accuracy of the sub-optimal approximation and highlight the interdependence between \( w_i \) and \( P_{\text{th}} \).}

		\label{sb4}
	\end{center}
\end{figure}

To provide further insight, in Fig. \ref{sb4}, we compare the sub-optimal value obtained from \eqref{sd5} with the exact value obtained through numerical solutions for two different values of \( P_{\text{th}} \). The results clearly demonstrate the accuracy of the sub-optimal approximation derived from \eqref{sd5}. 
Another important observation is that the optimal \( w_i \) varies as \( P_{\text{th}} \) changes. Previously, we showed that the optimal \( P_{\text{th}} \) also changes as \( w_i \) varies, highlighting the strong interdependence between these two parameters.

\begin{algorithm}[t]
	\caption{Iterative Joint Optimization of \( w_i \) and \( P_{\text{th}} \)}
	\label{alg:joint_opt}
	\begin{algorithmic}[1]
		\REQUIRE Instantaneous link length \( L_i \), background noise std.\ \(\sigma_{bg,i}\), tracking error std.\ \(\sigma_\theta\), transmitted power \( P_{t,i'} \), aperture radius \( r_a \), and an initial guess \(\bigl(P_{\text{th}}^{(0)}, w_i^{(0)}\bigr)\).
		
		\STATE \textbf{Initialize:}
		\STATE \hspace{\algorithmicindent} Set iteration counters \(n = 0\) (outer iteration) and \(\ell = 0\) (inner iteration).
		\STATE \hspace{\algorithmicindent} Choose a convergence tolerance \(\epsilon\) (e.g., \(10^{-3}\)) and maximum number of iterations \(N_{\max 1}\) and \(N_{\max 2}\) for inner and outer loops.
		
		\STATE \textbf{Outer Iteration:} 
		\REPEAT
		\STATE \textit{(1) Update \( w_i \) given \( P_{\text{th}} \):}
		\STATE \hspace{\algorithmicindent} Use \eqref{sd5} to compute a new \( w_i \):
		\begin{equation*}
			w_i^{(n+1)} = 
			f_{\text{beam}}\Bigl(P_{\text{th}}^{(n)}, L_i, \sigma_{\theta}, r_a, P_{t,i'}\Bigr),
		\end{equation*}
		where \( f_{\text{beam}} \) represents the closed-form expression in \eqref{sd5}.
		
		\STATE \textit{(2) Update \( P_{\text{th}} \) given \( w_i \) (Inner Iteration):}
		\STATE \hspace{\algorithmicindent} Set \(\ell = 0\) and initialize \( P_{\text{th}}^{(n,0)} = P_{\text{th}}^{(n)} \).
		\REPEAT
		\STATE \hspace{\algorithmicindent} Use \eqref{gr7} to refine \( P_{\text{th}} \): $P_{\text{th}}^{(n,\ell+1)}$=
		$f_{\text{threshold}}
			\Bigl(w_i^{(n+1)}, P_{\text{th}}^{(n,\ell)}, L_i, \sigma_{bg,i}, \sigma_{\theta}, P_{t,i'}, r_a\Bigr)$,
		where \( f_{\text{threshold}} \) corresponds to the expression in \eqref{gr7}.
		\STATE \hspace{\algorithmicindent} \(\ell \leftarrow \ell + 1\)
		\UNTIL convergence of \( P_{\text{th}}^{(n,\ell)} \) or \(\ell \geq N_{\max 1}\)
		\STATE \hspace{\algorithmicindent} Define \( P_{\text{th}}^{(n+1)} = P_{\text{th}}^{(n,\ell)} \).
		
		\STATE \textit{(3) Check Outer Convergence:}
		\STATE \hspace{\algorithmicindent} If 
		\(\bigl|P_{\text{th}}^{(n+1)} - P_{\text{th}}^{(n)}\bigr| < \epsilon \) 
		\textbf{and} 
		\(\bigl|w_i^{(n+1)} - w_i^{(n)}\bigr| < \epsilon\), 
		\textbf{then} \textbf{break}.
		\STATE \hspace{\algorithmicindent} \( n \leftarrow n + 1 \)
		
		\UNTIL \( n \geq N_{\max 2} \)
		
		\STATE \textbf{Output:} 
		\STATE \hspace{\algorithmicindent} \(\bigl(P_{\text{th}}^{*}, w_i^{*}\bigr) = \Bigl(P_{\text{th}}^{(n)}, w_i^{(n)}\Bigr)\).
		
	\end{algorithmic}
\end{algorithm}

Using the obtained results, we propose an iterative scheme in Algorithm~\ref{alg:joint_opt} that aims to balance two critical considerations in OHL-based multi-relay systems: achieving near-optimal performance and maintaining low computational complexity. A fully joint optimization of \( w_i \) and \( P_{\text{th}} \) would require substantial processing power and precise, real-time knowledge of channel conditions across multiple satellite nodes—an approach generally infeasible in the rapidly evolving environment of LEO. By contrast, our proposed method employs closed-form approximations from \eqref{sd5} and \eqref{gr7} to iteratively refine each parameter based on the latest updates of the other. This process is organized into an outer iteration, which alternates between updating the beam width \( w_i \) and refining \( P_{\text{th}} \) using an inner iteration. 
Following each update, the algorithm checks whether changes in both parameters remain below a preset tolerance, thus signaling convergence. This structure enables timely responses to fluctuations in link distance and tracking errors, as it can be re-invoked whenever the inter-satellite distance changes significantly or at fixed scheduling intervals. Simulation results  demonstrate that the method converges to near-optimal values quickly. Moreover, the overhead incurred by this iterative approach is substantially lower than that of a global solver, making it well-suited to real-time adaptation in LEO satellite networks where processing power and bandwidth are limited.

\begin{figure}
	\begin{center}
		\includegraphics[width=3.4 in]{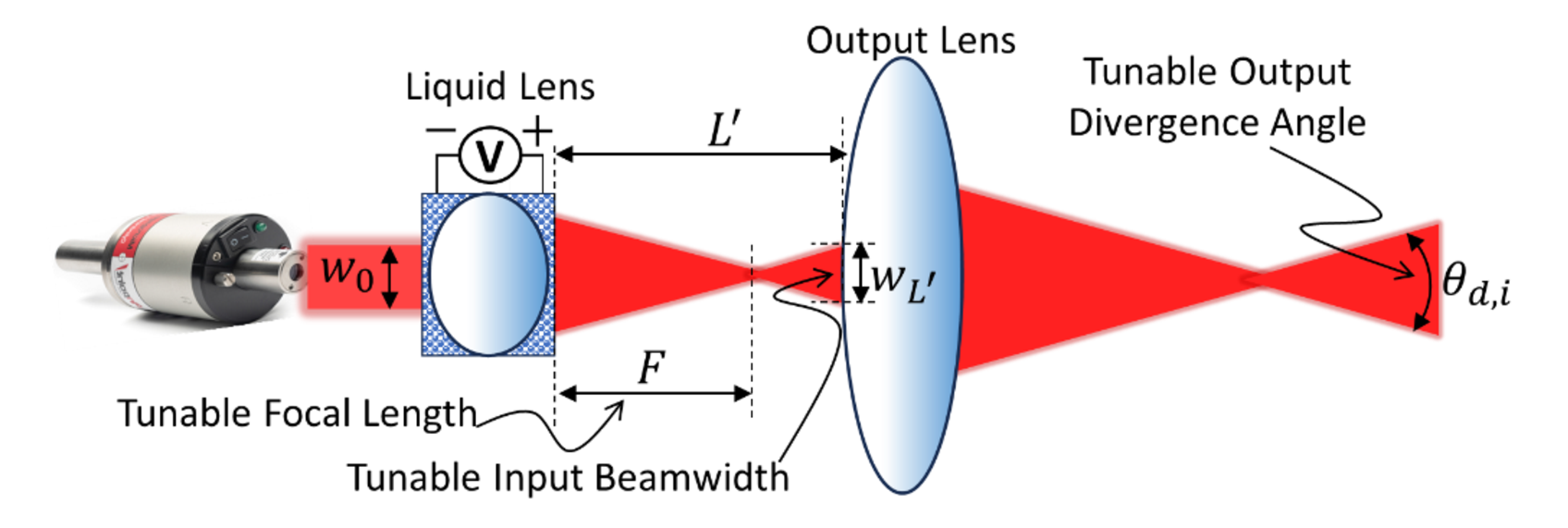}
		\caption{Schematic of the optical system with a tunable liquid lens. The system consists of a liquid lens with adjustable focal length \( F \), a fixed output lens at a distance \( L' \), and an incident Gaussian beam with initial beam width \( w_0 \). The liquid lens dynamically adjusts the beam waist, enabling control over the beam width \( w_{L'} \) and divergence angle at the output $\theta_{d,i}$.}
		\label{se2}
	\end{center}
\end{figure}

\section{Optimal Design of Beamwidth Control System}
This section presents a systematic approach for optimizing the beamwidth control system using tunable liquid lenses. We first review various beam divergence adjustment technologies and highlight the advantages of liquid lenses. Then, we describe the system architecture, derive the optimal focal length as a function of beam parameters, and discuss the voltage control mechanism required for real-time beamwidth adaptation.

\subsection{Technologies for Beamwidth Adjustment}
Several technologies can be utilized for beam divergence control, including mechanical zoom optics, adaptive mirrors, and deformable lenses. Mechanical zoom optics provide precise focal length adjustments but suffer from slow response times and increased system weight. Adaptive mirrors dynamically shape the outgoing wavefront but require complex control algorithms and high-precision actuators.

One of the emerging technologies that has recently gained significant attention and is rapidly evolving is liquid lens technology \cite{liu2023tunable}. Unlike traditional optics, liquid lenses can dynamically adjust the focal length $F_i$ by varying the applied voltage, thereby modifying the divergence angle $\theta_{d,i}$. This allows for precise control of beamwidth at the receiver to compensate for variations in link distance and alignment accuracy.

\subsection{System Overview}
Liquid lenses utilize \textit{electrowetting} or \textit{dielectric actuation} to adjust the curvature of the optical interface, enabling rapid modifications in focal length. 
The proposed optical system, as illustrated in Fig.~\ref{se2}, consists of the following key components:
\begin{itemize}
	\item A \textbf{liquid lens} with a dynamically tunable focal length \( F \), enabling real-time adjustment of the beam waist.
	\item A \textbf{fixed output lens} positioned at a distance \( L' \) from the liquid lens, which serves as the final optical element in the system.
	\item A Gaussian beam with an initial beam width \( w_0 \), which is incident on the liquid lens.
\end{itemize}
By dynamically adjusting the focal length \( F \) of the liquid lens through voltage variations applied across its electrodes, the beam waist is modified. This leads to a corresponding change in the beam width \( w_{L'} \) at the fixed output lens. The ability to control the focal length via voltage allows for precise and real-time adjustment of the beam divergence angle, making the system highly suitable for applications requiring adaptive beam control, such as inter-satellite communication in LEO.

\subsection{Optimal System Design}
In the following, we provide a detailed analysis of the relationship between the optimal beam width \( w_i^* \) at the receiver and the optimal focal length \( F \) of the liquid lens. By optimizing \( F \), we aim to achieve the desired \( w_i^* \) in real-time, ensuring efficient beam control in response to dynamic changes in the inter-satellite communication channel.

To achieve the optimal beam width \( w_i^* \) at the receiver for a given instantaneous link length \( L_i \), the divergence angle \( \theta \) at the transmitter must be adjusted according to the following relationship:
\begin{align} \label{sp7}
\theta_{d,i}^* = \frac{w_i^*}{L_i}.
\end{align}
This adjustment ensures that the beam remains optimally focused at the receiver, even as the link length \( L_i \) varies dynamically. By controlling the divergence angle through the tunable focal length \( F \) of the liquid lens, the system can maintain efficient power delivery and minimize beam spreading over the communication channel.
To achieve the desired optimal beam divergence angle \( \theta_{d,i}^* \) at the output of the fixed lens, the optimal input beam width \( w_{L'}^* \) must satisfy (see Fig. \ref{se2}):
\begin{align} \label{sp8}
	w_{L'}^* = \frac{\lambda}{\pi \theta_{d,i}^*}.
\end{align}

To achieve the desired beam width \( w_{L'}^* \) at the input of the fixed output lens, we must consider the effect of both the liquid lens and the propagation distance \( L' \). The liquid lens dynamically adjusts the beam waist, and after propagating a distance \( L' \), the beam width \( w_{L'}^* \) is both the output beam width of the liquid lens and the input beam width of the fixed output lens.
To determine the beam width at distance \( L' \) after passing through the liquid lens, we employ the ABCD matrix formalism. The total ABCD matrix for the system, consisting of a liquid lens with focal length \( F \) followed by free-space propagation over a distance \( L' \), is given by:
\begin{align}
	M = \begin{pmatrix}
		A & B \\
		C & D
	\end{pmatrix} = 
	\begin{pmatrix}
		1 & L' \\
		0 & 1
	\end{pmatrix}
	\begin{pmatrix}
		1 & 0 \\
		-\frac{1}{F} & 1
	\end{pmatrix}.
\end{align}
Multiplying these matrices, we obtain:
\begin{align} \label{sp2}
	M =
	\begin{pmatrix}
		1 - \frac{L'}{F} & L' \\
		-\frac{1}{F} & 1
	\end{pmatrix}.
\end{align}
The complex beam parameter \( q \) for a Gaussian beam is defined as:
\begin{equation}
	\frac{1}{q} = \frac{1}{R} - \frac{i \lambda}{\pi w^2}.
\end{equation}
where \( R \) is the radius of curvature of the wavefronts, \( w \) is the beam radius (spot size), and the imaginary part of \( \frac{1}{q} \), given by \( \frac{\lambda}{\pi w^2} \), represents the beam divergence and determines the Gaussian beam's width variation along propagation.

At the input (immediately before the liquid lens), the beam has a waist \( w_0 \), and its complex beam parameter is:
\begin{align}
	q_0 = i z_R = i \frac{\pi w_0^2}{\lambda}.
\end{align}
After passing through the liquid lens and propagating a distance \( L' \), the complex beam parameter \( q' \) transforms as:
\begin{align}
	q' = \frac{A q_0 + B}{C q_0 + D}. 
\end{align}
Using \eqref{sp2} and substituting \( q_0 = i z_R \), we obtain:
\begin{align} \label{sp4}
	q' = \frac{\left(1 - \frac{L'}{F}\right) i z_R + L'}{-\frac{1}{F} i z_R + 1}.
\end{align}
First, we rewrite \eqref{sp4} as
\begin{align}
	\frac{1}{q'} = \frac{\text{Re}(q') - i \text{Im}(q')}{|q'|^2}.
\end{align}
Substituting \( q' \) and simplifying, the imaginary part of \( \frac{1}{q'} \) is given by:
\begin{align} \label{sp5}
	\text{Im} \left( \frac{1}{q'} \right) = \frac{\lambda}{\pi w_{L'}^2} = \frac{z_R F}{\left(F - L'\right)^2 + z_R^2}.
\end{align}
Using \eqref{sp5} and after some manipulations, the optimal focal length \( F^* \) is derived as:
\begin{align} \label{sp9}
	&F^{*} = \\
	&\frac{ 2 L' + \frac{\pi (w_{L'}^*)^2 z_R}{\lambda} + \sqrt{\left( 2 L' + \frac{\pi (w_{L'}^*)^2 z_R}{\lambda} \right)^2 - 4 (L'^2 + z_R^2)} }{2}.  \nonumber
\end{align}
Finally, using \eqref{sp7}, \eqref{sp8}, and \eqref{sp9}, the optimal value of \( F^* \) as a function of the optimal beam width \( w_i^* \) is obtained in the form of \eqref{sp10}.
\begin{figure*}[!t]
	\normalsize
	\begin{align} \label{sp10}
		F^* = L' + \frac{\pi \left( \frac{\lambda L_i}{\pi w_i^*} \right)^2 z_R}{2\lambda} 
		+ \frac{1}{2} \sqrt{\left( 2 L' + \frac{\pi \left( \frac{\lambda L_i}{\pi w_i^*} \right)^2 z_R}{\lambda} \right)^2 - 4 (L'^2 + z_R^2)}.
	\end{align}
	\hrulefill
\end{figure*}

\subsection{Voltage Control and Response Time}
Once the optimal focal length $F^{\star}$ is determined, the liquid lens voltage is adjusted accordingly to achieve this value. The voltage-focal length mapping is typically characterized by a nonlinear function dependent on the liquid properties and electrode configuration.

Liquid lenses exhibit a rapid response time, typically in the range of:

\begin{align}
	t_{\text{response}} \approx 1 - 10 \text{ ms}.
\end{align}
This speed is sufficient for real-time adjustments in LEO satellite links, as the link length distribution and alignment conditions do not undergo significant changes within such short time intervals.

\section{Simulations and Discussions}

\begin{figure*}
	\centering
	\subfloat[] {\includegraphics[width=2.3 in]{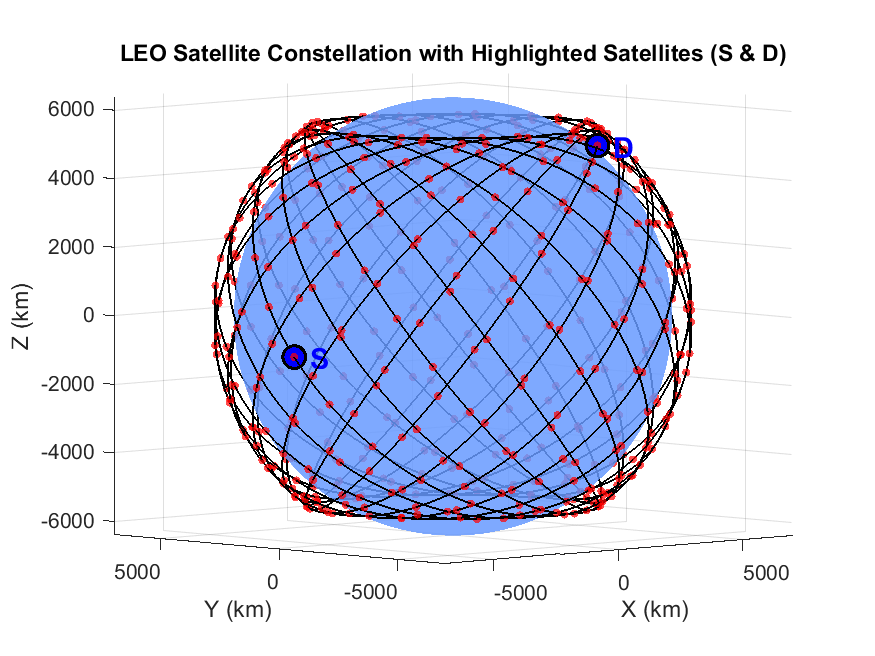}
		\label{b1}
	}
	\hfill
	\subfloat[] {\includegraphics[width=2.3 in]{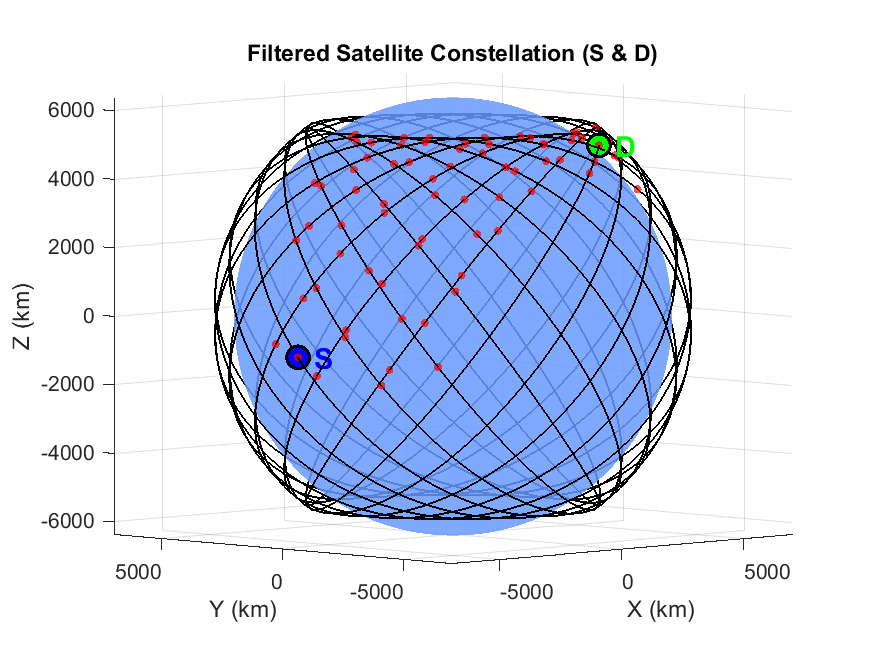}
		\label{b2}
	}
	\hfill
	\subfloat[] {\includegraphics[width=2.3 in]{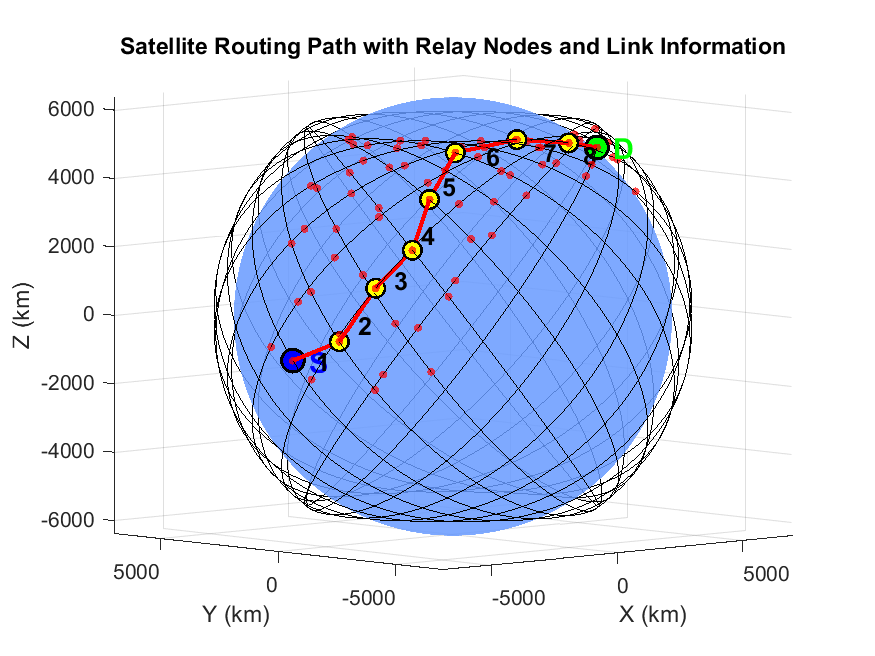}
		\label{b3}
	}
	\caption{An illustrative snapshot of the satellite constellation and relay path selection, where the network topology and routing continuously change due to satellite motion. (a) The full satellite constellation with multiple orbital planes and satellites. (b) A reduced set of candidate satellites selected to speed up the routing protocol by filtering out less probable relay nodes. (c) The final relay path obtained by applying the routing algorithm on the set of selected candidate satellites at this specific moment. This topology dynamically evolves as satellites move, requiring continuous adaptation of the relay path.}
	
	\label{b4}
\end{figure*}

\begin{table*}
	\caption{Optimized relay parameters for the selected inter-satellite relay path. The table presents the link types, link lengths, and the optimized parameters including the optimal decision threshold \( P_{\text{th}}^* \), beam width \( w_i^* \), and required focal length for each relay satellite. Additionally, the results obtained from the exhaustive search method and the corresponding error probabilities for both optimization approaches are provided. The comparison demonstrates the effectiveness of the proposed sub-optimal algorithm in achieving near-optimal performance with significantly reduced computational complexity.}
	\centering 
	\begin{tabular}{c } 
		\includegraphics[width=6.3 in]{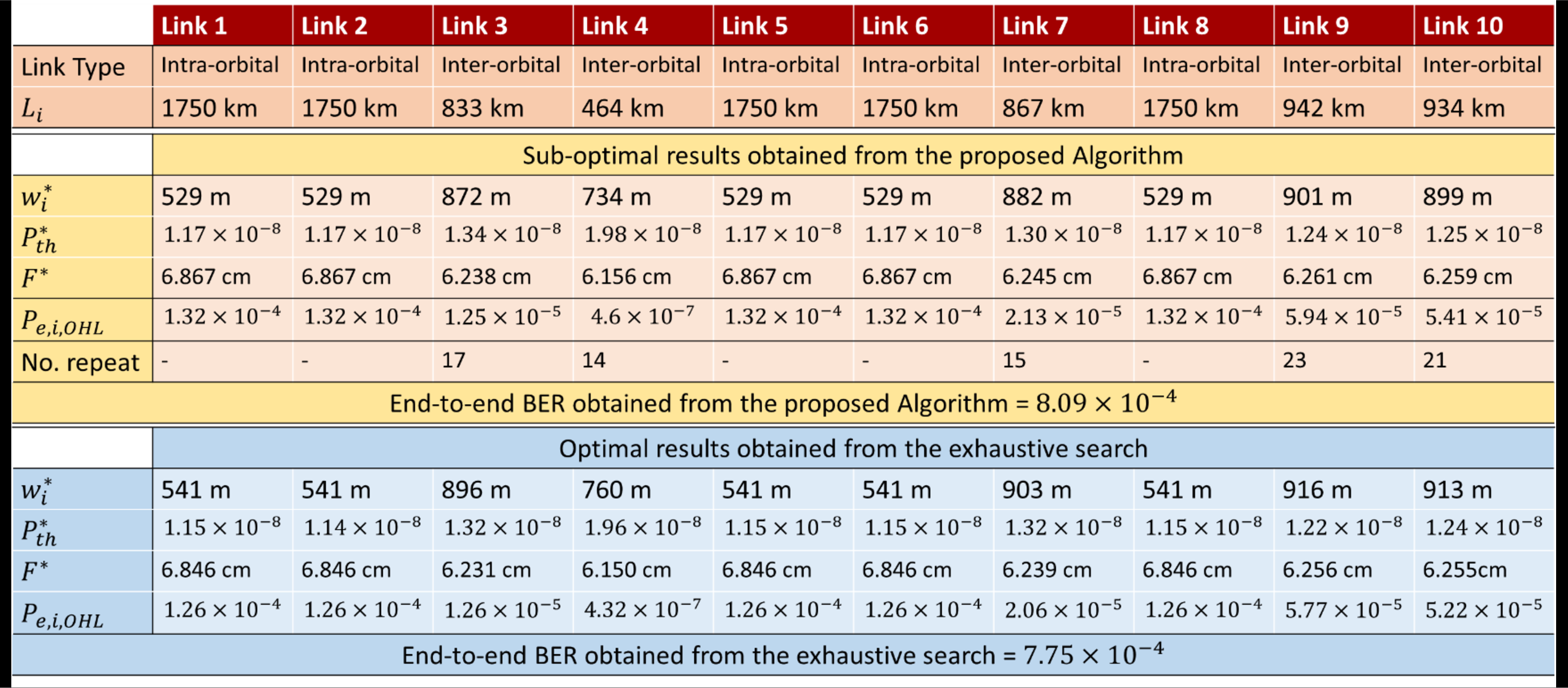}     
	\end{tabular}
	\label{sf2}
\end{table*}

To evaluate the performance of the OHL-based relay system, we consider two ground stations separated by a distance of 14,125 km. The objective is to establish a reliable communication link between these two points using a network of Low Earth Orbit (LEO) satellites. Since the technology used for satellite-to-ground links differs from inter-satellite optical links, we exclude the first and last hop, which correspond to the links between the ground station and the first satellite, and the last satellite and the destination. Instead, we focus on the multi-relay optical link established through multiple OHL-based relays.

To ensure a realistic simulation environment, we model a satellite constellation similar to the Starlink network, consisting of multiple orbital planes. Specifically, we generate 20 orbital planes, each containing 25 satellites, resulting in a total of 500 satellites in LEO. The satellites are placed at an altitude of 600 km, which is a typical operational altitude for modern LEO constellations. Each orbital plane is inclined at 53 degrees, closely resembling the design of large-scale LEO constellations aimed at providing global coverage.

The satellite positions are determined based on a uniform angular distribution within each orbit, ensuring consistent coverage. To emulate natural irregularities in satellite spacing due to deployment and orbital drift, each satellite’s position is perturbed by a small random angular offset within its orbital plane. Additionally, to determine the relay path, we identify two satellites that are closest to the given initial positions, ensuring that the communication path between them is realistic.

The LEO constellation used in the simulation is illustrated in Fig. \ref{b1}, where all the satellites and their orbital planes are depicted. This setup provides a practical foundation for evaluating the proposed OHL-based relay system under real-world conditions, accounting for key orbital parameters and inter-satellite distances. In the following sections, we present the simulation results and analyze the performance of the relay system.

For data relay from the source to the destination, two types of links are considered: intra-orbit and inter-orbit links. Intra-orbit links typically have a fixed distance between satellites, resulting in lower tracking system errors compared to inter-orbit links, where satellites move past each other at high relative velocities. Due to these dynamics, the tracking accuracy for inter-orbit links is set to 150~\micro rad, while for intra-orbit links, it is assumed to be \(\sigma_\theta=50\)~\micro rad. Given the system parameters, inter-orbit links are constrained to a maximum length of 1000~km in the routing protocol.

Since routing protocol design is beyond the scope of this paper, we use a near-realistic example by considering a snapshot of a specific relay path from source to destination. The resulting relay route for one instance is illustrated in Figures \ref{b2} and \ref{b3}.

After routing, the relay path consists of 10 relay links, whose lengths and link types (inter-orbital and intra-orbital) are provided in Table \ref{sf2}. Using this information, the proposed optimization algorithm rapidly computes the optimal parameters, including the optimal threshold \( P_{\text{th}}^* \) for each OHL, the optimal beam width \( w_i^* \) for each relay satellite, and the required focal length to achieve the optimal \( w_i^* \). These computed values are presented in Table \ref{sf2} to ensure a reliable relay link between the source and destination.

Additionally, results obtained from an exhaustive search-based simulation, which is significantly more time-consuming, are also included in the table. The error probabilities achieved using both the proposed sub-optimal algorithm and the exhaustive search method are compared. The results clearly demonstrate that the proposed sub-optimal algorithm not only achieves values close to the optimal ones but also ensures an error probability comparable to that of the ideal system optimized via exhaustive search. This validates the effectiveness of the proposed method in maintaining the quality and reliability of the relay link.


\begin{table}[h!]
	\centering 
	\caption{Average BER across multiple inter-satellite relay snapshots}
	\label{tab:multi_snapshot_results}
	\begin{tabular}{lrr}
		\hline
		Snapshot     & $N_r$ & E2E BER  \\
		\hline
		S1 (Table 2) & 10            & 0.00082                                 \\
		S2           & 12            & 0.00065                                  \\
		S3           & 10            & 0.00091                                 \\
		S4           & 11            & 0.00068                                  \\
		\hline
	\end{tabular} 
\end{table}

It is important to note that the results reported in Table~\ref{tab:multi_snapshot_results} represent only one snapshot of an inter-satellite relay system at a specific instant. Given the high velocities and continuous movement of satellites, the topology and positions of relay nodes change dynamically over time. Consequently, optimization must be performed repeatedly and in real-time.
To provide a clearer understanding of the impact of network dynamics over long distances, four instantaneous snapshots of the described inter-satellite topology—spanning a total source-to-destination distance of 14{,}125~km—were captured. For each snapshot, a distinct relay path was identified, comprising 10, 12, 10, and 11 relay satellites, respectively. By applying the proposed optimization algorithm, it was observed that the average end-to-end BER for the optimal exhaustive search and the proposed algorithm were closely matched, as summarized in Table~\ref{tab:multi_snapshot_results}.
The purpose of presenting these results is to demonstrate that the network topology is inherently dynamic due to the high mobility of satellites. Consequently, the optimal parameters must be updated frequently in real-time to maintain robust system performance. The proposed algorithm effectively adapts to these dynamic conditions, ensuring reliable operation of the all-optical inter-satellite relay system based on OHL technology.

\begin{table*}[t] 
	\caption{Comparison of Inter-Satellite Relay Techniques}
	\label{tab:relay_comparison}
	\centering
	\renewcommand{\arraystretch}{1.3}
	\begin{tabular}{@{}l|cccccccc@{}}
		\toprule
		\textbf{Method / Feature} & 
		\makecell{O/E and \\ E/O Conversion} & 
		\makecell{Processing \\ Delay} & 
		\makecell{Processing \\ Complexity} & 
		\makecell{Noise \\ Accumulation} & 
		\makecell{Threshold \\ Sensitivity} & 
		\makecell{Channel \\ Estimation} & 
		\makecell{Robustness to \\ Channel Variation} & 
		\makecell{Power \\ Consumption} \\
		\midrule
		DF & Yes & Yes & Yes & No  & No  & Yes & High   & High \\
		AF & No  & No  & Low & Yes & No  & No  & Low & Lower \\
		OHL (Proposed) & No  & No  & Very Low & No  & Yes & No  & Medium* & Lower \\
		\bottomrule
	\end{tabular}
	\begin{flushleft}
		\footnotesize
		*Robustness in OHL are design-dependent and can be optimized through threshold control and passive components.
	\end{flushleft} 
\end{table*}

The overall performance and implementation comparison of DF, AF, and the proposed OHL-based relay is summarized in Table~\ref{tab:relay_comparison}. As evident, the OHL method offers a uniquely favorable trade-off: it eliminates the need for optical-electrical conversions, exhibits minimal processing delay and complexity, and avoids noise amplification. Although its robustness depends on optimal threshold design, this can be efficiently managed using passive photonic components. Therefore, beyond its competitive performance in BER and outage simulations, the OHL-based scheme stands out as a practical and energy-efficient solution for fully optical inter-satellite communication systems.

\section{Conclusion and Future Directions} 
In this work, we presented a comprehensive analysis of inter-satellite relaying based on Optical Hard Limiters (OHL), aiming to minimize latency while maintaining high reliability in Low Earth Orbit (LEO) constellations. By jointly optimizing the OHL threshold and beam divergence angle—and integrating tunable liquid lenses for real-time beamwidth control—we demonstrated substantial gains in error performance and power efficiency compared to conventional Amplify-and-Forward (AF) and Decode-and-Forward (DF) approaches. Our simulation results, carried out in a large-scale LEO constellation scenario, showed that the proposed sub-optimal algorithm can achieve near-optimal parameter settings with significantly reduced computational overhead, making it suitable for dynamic environments where link distances and network topology continuously change.

\begin{itemize}
	\item \textit{Routing and resource allocation:} Incorporating route selection into the design process so that relay paths and OHL parameters are jointly optimized, potentially reducing end-to-end latency further in large constellations.
	\item \textit{Multi-threshold OHL relays:} Exploring the possibility of adaptive or hybrid threshold switching at each relay to enhance robustness under rapidly varying channel conditions.
	\item \textit{Machine-learning-based optimization:} Employing reinforcement learning or other data-driven approaches to handle limited or stochastic channel feedback, enabling on-the-fly, near-optimal parameter adaptation in dynamic satellite networks.
	\item \textit{Experimental validation:} Extending this work through hardware-in-the-loop or in-orbit demonstrations to validate mechanical stability, beam pointing accuracy, and liquid lens performance, ultimately facilitating practical deployment of OHL-based inter-satellite relaying in next-generation networks.
\end{itemize}
By addressing these directions, future research can further solidify the role of OHL-based relays in next-generation satellite networks, enabling higher data rates, greater coverage, and truly global low-latency connectivity.

\balance

\end{document}